\documentstyle[aps,12pt,prb]{revtex}

\begin{document}
\title{New criteria for cluster identification in continuum systems.}
\author{Luis A. Pugnaloni and Fernando Vericat$^{*}$.}
\address{{\it Instituto de F\'{i}sica de L\'{i}quidos y Sistemas Biol\'{o}gicos }\\
{\it (IFLYSIB)-UNLP-CONICET cc. 565 - (1900) La Plata, Argentina}\\
{\it and }\\
{\it Grupo de Aplicaciones Matem\'{a}ticas y Estad\'{\i }sticas de la}\\
Facultad de Ingenier\'{\i }a (GAMEFI) - Departamento de\\
Fisicomatem\'{a}tica, Facultad de Ingenier\'{i}a. Universidad Nacional de La%
\\
Plata, La Plata, Argentina}
\maketitle

\begin{abstract}
\bigskip

Two new criteria, that involve the microscopic dynamics of the system, are
proposed for the identification of clusters in continuum systems. The first
one considers a residence time in the definition of the bond between pairs
of particles, whereas the second one uses a life time in the definition of
an aggregate. Because of the qualitative features of the clusters yielded by
the criteria we call them chemical and physical clusters, respectively.
Molecular dynamics results for a Lennard-Jones system and general
connectivity theories are presented.


$^{*}$ Correponding author
\end{abstract}

\section{Introduction}

Clustering plays a relevant role in condensed matter physics, particularly
in relation with transition phenomena, as it has been recognized since more
than sixty years ago when Bijl, Band and Frenkel (among others)\cite
{Bijl1,Band1,Frenkel1} introduced the concept of real clusters, in place of
the mathematical ones that Mayer had considered in the virial expansion of
imperfect gases\cite{Mayer1}.

The application of statistical mechanics formalism to describe clustering in
equilibrium classical systems is mainly due to Hill\cite{Hill1}. In Hill's
theory the concept of cluster is directly related to that of connectivity:
two particles belong to the same cluster if they are connected through a
path of directly connected particles. Therefore a crucial point in the
identification of the clusters is the definition of directly connected
particles (a bonded pair). Whereas thermodynamic properties are not affected
by the particular definition used to identify the clusters, clustering and
percolation properties do are very sensitive to this choice instead\cite
{Hill1}.

Hill's definition of a bonded pair is based on an energetic criterion: two
particles are bonded, in a given configuration, if their relative kinetic
energy is less than minus the pair potential energy\cite{Hill1}. Frequently,
however, a geometrical criterion is used instead of the energetic one. Thus,
Stillinger criterion states that two particles are bonded if they are
separated by a distance shorter than certain value $d$, which is called the
connectivity distance\cite{Stillinger1}.

In the past few years, several papers have renewed interest in the energetic
and dynamic aspects of bonding, particularly in relation to the properties
of water\cite{Sciortino1}. Hydrogen-bond life times have been calculated and
new connectivity criteria have been proposed for water molecules. These
works suggest that a bonded pair must be defined as two water molecules that
are in some appropriate geometrical arrangement at least during a time
interval of the order of the estimated hydrogen-bond life times. Recently
has been pointed out the necessity of including dynamical information in the
definition of the clusters considered in nucleation studies\cite
{Senger1,Schenter1,Kusaka1}. It is also expected that taking into
consideration the dynamics of aggregates would improve the description of
some phenomena as the conductor/insulator transitions in microemulsions\cite
{Cametti1}.

Hill's theory separates the Boltzmann factor $e({\bf r}_{1},{\bf r}%
_{2})=\exp [-\beta v({\bf r}_{1},{\bf r}_{2})]$ into bonded $(\dagger )$ and
unbounded $(*)$ terms: $e({\bf r}_{1},{\bf r}_{2})=e^{\dagger }({\bf r}_{1},%
{\bf r}_{2})+e^{*}({\bf r}_{1},{\bf r}_{2})$. In this paper we will consider
only systems interacting via pairwise additive potentials and will denote
with $v({\bf r}_{1},{\bf r}_{2})$ the pair potential. As usual $\beta $ $%
=1/k_{B}T$ with $T$ the absolute temperature and $k_{B}$ the Boltzmann
constant. Since $e^{\dagger }({\bf r}_{1},{\bf r}_{2})$ represents the basic
probability density that two particles at positions ${\bf r}_{1}$and ${\bf r}%
_{2}$, respectively, are directly connected (bonded), this separation gets a
diagrammatic expansion for the partition functions in terms of real clusters
instead of the mathematical clusters of Mayer..

Fugacity and density expansions similar to those obtained by Mayer and
Montroll\cite{Mayer2} for the ordinary pair correlation function $g({\bf r}%
_{1},{\bf r}_{2})$ has been found, within Hill's formalism, by Coniglio and
collaborators for the pair connectedness function $g^{\dagger }({\bf r}_{1},%
{\bf r}_{2})$. This function is proportional to the joint probability
density of finding two particles connected (directly or indirectly) and at
positions ${\bf r}_{1}$ and ${\bf r}_{2}$, respectively\cite{Coniglio1}.
Moreover, by collecting nodal and non-nodal diagrams in these expansions an
Ornstein-Zernike like relationship is obtained. Consequently, integral
equations for $g^{\dagger }({\bf r}_{1},{\bf r}_{2})$ can be posed\cite
{Stell1}. These equations, as well as computer simulations, have been
applied together with Stillinger criterion to study clustering and
percolation in several continuum systems. They include: the ideal gas\cite
{Chiew1,Chiew2,Gawlinski1,Sevick1}; simple hard spheres\cite{DeSimone1,Lee1}%
; hard spheres with (sticky) adhesion\cite{Chiew1,Seaton1}; hard spheres
with square wells\cite{Netemeyer1,Chiew3}; hard spheres with Yukawa tails%
\cite{Xu1}; charged hard spheres\cite{Bresme1}; Lennard-Jones fluids\cite
{Heyes1} and dipolar hard spheres\cite{Vericat1,Laria1,Carlevaro1}.

Hill's work is the basis for all the present clustering theories in
continuum systems. Nevertheless, it has some limitations. In one hand, as
Hill pointed out in his paper\cite{Hill1}, the essence of his theory is the
concept of bonded pair. This fact limits the kind of cluster we can study.
In the other hand both, the energetic and the geometrical, criteria
considered by Hill yield a poor description of a ``real'' bond.

Let first consider the second problem. If we use an energetic criterion to
define a bond, then the functions $e^{\dagger }$ and $e^{*}$ must be
momenta, as well as position, dependent. However, the functions $e^{\dagger
} $ and $e^{*}$ used by Hill are only position dependent, since the momenta
have been averaged. A geometrical criterion is neither enough by itself to
warrant the existence of a real bond. After a short time two neighbor
particles can be much separated each other if the relative velocity is high.
These features suggest that the bond definition should be modified. In
Section II we introduce a bond criterion which incorporates a residence time
into Stillinger criterion. We shall call ``chemical clusters'' the
aggregates so defined. A connectivity theory, already presented in a
previous paper\cite{Pugnaloni1}, that allows to handle these clusters will
be reviewed in Subsection V-A.

Let us now focus on the energetic criterion and take, for example, three
particles which remain next one to the others. If the potential interaction
between each pair is not strong enough to form a bond then, according to
Hill's theory, we will have three clusters of one particle each.
Nevertheless, since the three particles remain close each other, then the
system can be considered as a single cluster of three particles
``collectively or non-specifically pair-bonded''. We call this kind of
aggregates ``physical clusters''.

With respect to the geometrical criterion, it could happen that it does not
matter how the particles interact each other, if they are close enough, then
they are bonded. Thus, according to the Stillinger's criterion, there will
be three bonded pairs and the particles will form a single three-particle
cluster. In this way the Stillinger's criterion partly avoid the limitation
that the energetic criterion imposes in Hill's work. Due to this feature,
Stillinger's criterion has given good results in nucleation studies\cite
{Senger1}. However, it does not include dynamical information. As a
consequence one could take as a single cluster two physical clusters which
are close during a very short time. In Section III we will give a definition
for physical clusters which adds a dynamical restriction to the Stillinger
criterion. This restriction is equivalent to require a minimum lifetime for
the aggregates. In Subsection V-B we present a simplified theory to obtain
connectivity functions and mean cluster sizes for physical clusters.

It can be worth-mentioning that the difference between the chemical and
physical clusters that we introduce is, in some sense, analogous to that
existing between chemical (or specific) and physical (or non-specific)
adsorption.

We remark that in this work, the time is thought in the frame of the
microscopical evolution of the system whereas its macroscopic
(thermodynamic) state remains unchanged.

Since the work of Coniglio and collaborators, the central object in
clustering and percolation theory of continuum systems is the cluster pair
correlation function $g_{\gamma }({\bf r}_{1},{\bf r}_{2})$. It is the joint
probability density of finding two particles that belong to the same cluster
of kind $\gamma $ ($\gamma =$Stillinger, chemical or physical) at positions $%
{\bf r}_{1}$and ${\bf r}_{2}$, respectively. In particular, for Stillinger
clusters we have $g_{Still}({\bf r}_{1},{\bf r}_{2})=g^{\dagger }({\bf r}%
_{1},{\bf r}_{2})$. From cluster correlation functions we can calculate the
mean cluster size defined

\begin{equation}
S_{\gamma }(\rho )=1+\frac{1}{(N-1)}\int g_{\gamma }({\bf r}_{1},{\bf r}%
_{2})d{\bf r}_{1}d{\bf r}_{2},  \label{n1}
\end{equation}
where $N$ is the number of particles of the system and the integrations are
over the whole system volume $V.$ In Eq.(\ref{n1}) we have explicitly
indicated its dependence on the system density $\rho =N/V$. Percolation of
an ensemble of particles is concerned with the existence of clusters that
become macroscopic in size. Then the percolation transition occurs at a
critical percolation density $\rho _{c}$ which mathematically verifies

\begin{equation}
\lim\limits_{\rho \rightarrow \rho _{c}^{-}}S_{\gamma }(\rho )=\infty .
\label{n2}
\end{equation}

The rest of this article is arranged in the following way. In Section II and
III we introduce the chemical and physical clusters, respectively. Section
IV will be devoted to discuss some results from molecular dynamics
simulations. In section V we present statistical mechanics theories to
describe both kind of clusters. We conclude with some comments on potential
applications of these new criteria.

\section{Chemical clusters}

As it was mentioned, Stillinger's criterion does not describe a real bond in
a completely satisfactory way since it does not include any dynamical
information. For example, two particles which are separated by a distance
shorter than $d$, at a given instant, are considered bonded even when their
relative velocity is high enough to separate the particles in a short time.
On the other hand, the analysis of the residence time of a molecule in the
neighborhood of another one has produced important advances in the study of
the hydrogen bond in water\cite
{Sciortino1,Sciortino2,Abraham1,Luzar1,Harrington1,Starr1}. All those facts
suggest to include a residence time to describe a real bond. In this
direction we propose the following bond definition:

\begin{center}
{\bf Two particles in the system are bonded at the instant }$t${\bf \ if
they have been separated by a distance shorter than the connectivity
distance }$d${\bf \ during the time interval [}$t-\tau ,t${\bf ], where }$%
\tau $\ {\bf is the residence time. }
\end{center}

Then we define a cluster in the same way as Hill did it:

\begin{center}
{\bf A chemical cluster is a set of particles such that between any pair of
them exists a path of bonded particles.}
\end{center}

As we can see, the residence time modifies the bond definition but, once the
definition of a bond is given, the same connectivity rule for the cluster
identification is used. We call chemical clusters to this aggregates because
the stable bonds are the essence of the cluster formation, like the covalent
bond is the essence in the building of molecules\cite{Nota1}. In the limit
case $\tau =0$ we recover Stillinger's definition.

The numbers $d$ and $\tau $ are two parameters of the theory that must be
obtained from physical considerations on the system and the phenomenon under
study. For example, in nucleation studies for Lennard-Jones fluids, where
the particles interact via the pair potential

\begin{equation}
v({\bf r}_{i},{\bf r}_{j})=v(r_{i,j})=4\varepsilon \left[ \left( \frac{%
\sigma }{r_{i,j}}\right) ^{12}-\left( \frac{\sigma }{r_{i,j}}\right)
^{6}\right] ,  \label{n3}
\end{equation}
is generally adopted the value $d=1.5\sigma $\cite{Senger2}. According with
its role in the theory, one possible election of $\tau $ could be done on
the basis of considering the mean value of the relative velocity between two
atoms in the fluid at temperature $T$:

\[
\overline{v_{1,2}}=\sqrt{\frac{6k_{B}T}{m}} 
\]
where $m$ denotes the mass of the atoms. Thus we choose $\tau $ as the
necessary time for the particles travel, in average, the connectivity
distance: 
\[
\tau =\frac{d}{\overline{v_{1,2}}}=d\sqrt{\frac{m}{6k_{B}T}}. 
\]

To have an idea of magnitude orders, let consider argon at $T=300$ K. We
have $\sigma =3.41\stackrel{\circ }{\text{A}}$\ , $\varepsilon =1.65\times
10^{-21}$J and $m=6.62\times 10^{-26}$ Kg, so that $d=5.03$ $\stackrel{\circ 
}{\text{A}}$\ and $\tau =0.83$ ps or, in reduced units, $\tau ^{*}\equiv
\tau \sigma ^{-1}\sqrt{\varepsilon /m}\cong 0.39$; $d^{*}=d/\sigma =1.5$.

It should be remarked that this estimation for $\tau $ is just an example.
The residence time will depend on the system and the particular phenomenon
to be studied and can be determined by very different causes. In section VI
we briefly comment on the application of chemical clusters to the case of
water in oil microemulsions. There we discuss in particular on the
determination of the parameters.

\section{Physical clusters}

Here, we will consider how to identify physical clusters, i.e. those
aggregates in which specific bonds between pairs of particles are not
necessary for the clusters existence. As we have mentioned, this kind of
clusters could be of interest in the simulation of nucleation process. The
Stillinger's criterion and some other improved criteria are currently used
in those studies\cite{Senger1}.

We propose, in order to define physical clusters, a criterion in which the
dynamics of the system is implicitly involved:

\begin{center}
{\bf A set of particles form a physical cluster in the instant }$t${\bf \ if
all they remained connected during the whole time interval [}$t-\tau ,${\bf %
\ }$t$]{\bf .}
\end{center}

Here the term ``connected'' must be understood in the Hill-Stillinger sense,
i.e. two particles are connected if between them there exists a path of
directly connected particles. Two particles being directly connected if they
are at a distance shorter than $d$.

The previous definition is incomplete. It must be accompanied by the
additional condition that those particles of the system which do not belong
to a given cluster can not contribute to the connectivity of that cluster.
Actually, the best way of stating the criterion is operationally. In this
vein it is convenient to introduce the concept of fragmentation event. A
fragmentation event occurs when a cluster breaks up into two or more
connected subaggregates.

Suppose we identify the system clusters at $t-\tau $ according to the (
``instantaneous'') Stillinger's criterion. Select one of these clusters.
When the system evolves from $t-\tau $ to $t$, their particles eventually
connect and disconnect several times. If we consider just those particles
which had belonged to the selected Stillinger cluster at $t-\tau $ and
consider the successive fragmentation events then, after the period $[t-\tau
,t]$, the particles will lie in several subaggregates. Each subaggregate
forms a physical cluster at time $t$.

In Fig. 1 we show a simple example where the differences among Stillinger,
chemical and physical clusters can be appreciated. At the initial time $%
t-\tau $ we show the instantaneous Stillinger clusters (a). At the final
time $t$ we identify Stillinger (b), physical (c) and chemical clusters (d).
We can see that physical clusters are smaller in size and longer in life
than Stillinger ones, which are instantaneous. Thus, physical clusters would
be more stable entities than Stillinger ones. However, physical clusters do
not need pair-bonded particles to exist whereas these last are essential for
defining chemical clusters. In Fig. 1, particles 1 and 3 form a physical
cluster of two particles, but they do not form a chemical cluster: although
they are directly connected at $t$ they were not at $t-\tau $.

An estimation of the characteristic time $\tau $ and the connectivity
distance $d$ is necessary for physical clusters as well as for chemical
clusters. In our calculations on Lennard-Jones systems, we use the same
value of $d$ that for the chemical clusters. In order to select a value for $%
\tau $ it is necessary to estimate the mean life time of an aggregate.
Nevertheless, we are interested in comparing the results of the two
different criteria. Thus, we will use the same value of $\tau $ for both,
chemical and physical criteria.

\section{Molecular dynamics}

\subsection{Chemical clusters}

In order to show the implications of using the residence time restriction in
the bond definition we have simulated by molecular dynamics (MD) a system of
particles which interact through a Lennard-Jones potential. Monte Carlo
simulation can not be used because the trajectory of the particles is
required.

We use a leap-frog algorithm with velocity correction to simulate a system
at constant volume and temperature\cite{Allen1}. To identify the chemical
clusters we modify the standard MD code in such a way that the resulting
algorithm is as follow:

{\bf 1)} All initial conditions are set for the MD ($t=t_{0}$).

{\bf 2)} The NEIGHBORS routine is called. It generates a table with all
direct connections -in the Stillinger sense- between the atoms. The
positions of the particles at $t=t_{0}$ are used in that operation.

{\bf 3)} The MD algorithm is carried out with a very short time step $\Delta
t$ until a pair of atoms disconnect in the Stillinger sense ($t=t_{1}$).

{\bf 4)} The REDUCTION routine is called. It checks and deletes from the
neighbors list the direct connections that have disappeared. The position of
the particles at $t=t_{1}$ are used in this step.

{\bf 5)} If $t_{1}-t_{0}<\tau $ go back to step {\bf 3}, else go to step 
{\bf 6.}

{\bf 6)} All clusters are separated by using the remaining neighbors list.
At this point the neighbors list only contain the direct connections which
survived a time $\tau $ from $t=t_{0}$, e.g. the list of chemical bonds.

{\bf 7)} All interesting information is saved (coordinates, cluster list,
etc.)

{\bf 8)} Initial conditions are set from the last configuration. Return to
step {\bf 2}.

Steps {\bf 1} to {\bf 8} are the loop for the chemical cluster
identification corresponding to the final configuration at time $t=t_{1}$.

In step {\bf 2} NEIGHBORS routine makes a list with the same structure of
the Verlet's neighbors list\cite{Allen1}, but a cut off equal to the
connectivity distance $d$ is used.

In step {\bf 3} the leap-frog algorithm is carried out with a time step $%
\Delta t$ short enough as to allow that no more than one disconnection
occurs. We chose $\Delta t$ to find a disconnection in each $3\Delta t$
intervals in average.

In step {\bf 4} the neighbors list created in {\bf 2} is reduced by deleting
the disappeared connections. The new connections created in the system are
discarded. Because in the last $\Delta t$ interval more than one
disconnection can occur, we check all connections in step {\bf 4} and any
disappeared connection is deleted from the neighbor list.

After the time interval $t_{1}-t_{0}$ grows (by successive iterations)
beyond $\tau $, a list of chemical bonds (as defined in section II) remains
in the neighbors list. Any standard algorithm for cluster identification can
be used to separate each cluster knowing the bonded pairs\cite
{Seaton1,Stoddard1,Hoshen1}. We have used Stoddard's one\cite
{Allen1,Stoddard1}.

\subsection{Physical clusters}

For physical clusters we use a similar method to that applied in the case of
chemical cluster. Nevertheless, the clustering steps are modified to achieve
the desired definition. The complete algorithm is:

{\bf 1)} All initial conditions are set for the MD ($t=t_{0}$).

{\bf 2)} Stillinger clusters are tabulated by standard routines\cite{Allen1}%
. The positions of the particles at $t=t_{0}$ are used in that operation.

{\bf 3)} The MD algorithm is carried out in very short time steps $\Delta t$
until a pair of atoms disconnect in the Stillinger sense ($t=t_{1}$).

{\bf 4)} The DIVISION routine is called. It checks each cluster in the table
made in step {\bf 2} and breaks up those clusters which were fragmented as a
result of the disconnection in step {\bf 3}. The cluster table is updated.

{\bf 5)} If $t_{1}-t_{0}<\tau $ go back to step {\bf 3}, else go to step 
{\bf 6.}

{\bf 6)} Save the interesting information. The cluster table contains all
the set of particles which are in agreement with the physical cluster
definition of section III.

{\bf 7)} Initial conditions are set from the last configuration. Return to
step {\bf 2}.

Steps {\bf 1} to {\bf 7} are the loop to identify the physical clusters for
the final configuration at time $t=t_{1}.$

In step {\bf 4} a clustering count must be done separately over each initial
cluster identified in {\bf 2. }In this way the connections formed between
different clusters in the time interval [$t_{0},t_{1}$] are not included in
the count.

\subsection{Results}

In this section we show the results obtained by molecular dynamics
simulation for a Lennard-Jones fluid with the pair potential given by Eq.(%
\ref{n3}). Our aim is not to exhaust the study of the various aspects of the
clustering that we will consider but instead is simply to remark the
qualitative differences they show for the three kinds of clusters:
Stillinger, chemical and physical.

We will express all quantities in reduced units: $\rho ^{*}=N\sigma ^{3}/V$, 
$T^{*}=k_{B}T/\varepsilon $, $\tau ^{*}=\tau \sigma ^{-1}\sqrt{\varepsilon /m%
}$ and $d^{*}=d/\sigma $ for density, temperature, residence and life time,
and connectivity distance respectively. We have used the leap frog algorithm
with velocity corrections at half step to produce a molecular dynamics with
constant number of particles $N$, volume $V$ and temperature $T$\cite{Allen1}%
. $500$ particles were placed in a cubic box with periodic boundary
conditions and $1000$ configurations were generated after stabilization to
calculate time averages. We use a cut off equal to $2.7\sigma $ for the pair
potential interaction and a tail correction for the remaining interaction.
Because we use the same number of particles in all simulations, scaling laws
were not used to extrapolate in the limit $N\rightarrow \infty $. The time
step is varied between $\Delta t^{*}=0.00001$ and $\Delta t^{*}=0.0001$
depending on temperature to get a connection or disconnection approximately
every three time steps. The parameters for the cluster identifications were
chosen $d^{*}=1.5$ and $\tau ^{*}=0.5$ for chemical and physical clusters.
In order to compare the results with the standard Stillinger criterion we
use the same connectivity distance $d^{*}=1.5$ for the three criteria.

In Fig. 2 the correlation functions for Stillinger clusters $%
g_{Still}(r_{1,2})$, chemical clusters $g_{Chem}(r_{1,2})$ and physical
clusters $g_{Phys}(r_{1,2})$ are shown. We have used free boundary
conditions in the calculation of the correlation functions for clusters\cite
{Lee1}. It means that neither real position nor nearest image of a particle
but the image belonging to the proper aggregate through the replicas is used.

If two particles are separated by a distance shorter than the connectivity
distance, then they belong to the same Stillinger cluster with certainty.
Thus, for $r_{1,2}<d$, the function $g_{Still}(r_{1,2})$ coincides with the
ordinary radial distribution function $g(r_{1,2})$ as it can be appreciated
in Fig. 2. The observed discontinuity at $r_{1,2}=d$ for all cluster
definitions is typical: the probability for two particles to be connected at 
$r_{1,2}>$ $d$, even for $r_{1,2}\rightarrow d^{+}$, depends on the presence
of a intermediate third particle and thus the probability of belonging to
the cluster notably disminishes. The function $g_{Still}(r_{1,2})$ is long
ranged because the density of the system considered in the figure coincides
with the percolation density for Stillinger clusters at the given
temperature. The mean Stillinger cluster size $S_{Still}$ diverges at this
density.

As it is expected $g_{Chem}(r_{1,2})$ and $g_{Phys}(r_{1,2})$ are smaller
than $g_{Still}(r_{1,2})$ for all $r_{1,2}$ because a dynamical restriction
is required in addition to the geometrical one. Comparing $g_{Chem}(r_{1,2})$
against $g_{Phys}(r_{1,2})$ in Fig. 2, we see that $g_{Chem}(r_{1,2})$ is
above $g_{Phys}(r_{1,2})$ for every $r_{1,2}$. That is just what one expect
when the same value for $\tau $ is used in both, chemical and physical
clusters definition, because a life time restriction on the aggregate is a
condition weaker than imposing a residence time restriction on the bond.

In Fig. 3 we present the system coexistence curve obtained by Panagiotopoulos%
\cite{Panagio1} together with the percolation curves for the various cluster
definitions. These curves separate the phase diagram in two parts:
percolated (high densities) and non-percolated (low densities); and
intersect the coexistence curve.

To determine if a given configuration is percolated we look for a cluster
which span through the replicas. For a given temperature we increase the
density until get 50\% of percolated configurations. In that case we say
that the percolation density was reached\cite{Seaton1}.{\it \ }We observe
that the percolation curve for physical clusters is slightly less steep than
for the Stillinger ones. Nevertheless, it is at higher densities as is
expected.

In the other hand, the percolation curve for chemical clusters is at higher
densities than for the physical ones and has a pronounced temperature
dependence. We will discuss in section VI how this temperature dependence of
chemical clusters makes them special candidates to describe the
metal-insulator transition in water in oil microemulsions.

Fig. 4 shows the size distribution function for Stillinger clusters $%
n_{Still}(s)$, chemical clusters $n_{Chem}(s)$ and physical clusters $%
n_{Phys}(s)$. These functions measure the number of clusters of size $s$ per
particle. For both, chemical and physical clusters, the number of small
clusters is larger than for the Stillinger ones. In the displayed example, $%
T^{*}=1.4$ and $\rho ^{*}=0.155$, the biggest chemical clusters are
approximately of size 60, the biggest physical clusters are of size 200 and
the biggest Stillinger clusters are of size 400. The life time restriction
in the physical clusters avoid the unphysical big aggregates yielded by the
Stillinger criterion. This fact suggest that physical clusters could be
useful in nucleation studies of saturated vapors.

In Fig. 5 the curves $n_{Still}(s)$, $n_{Chem}(s)$ and $n_{Phys}(s)$ are
shown at the corresponding percolation density for each cluster definition.
At the percolation density we obtain the power law\cite{Heyes1,Stauffer}: 
\begin{equation}
n(s)=As^{-z},  \label{n4}
\end{equation}
where the exponent has an universal value\cite{Stauffer}. Heyes and Melrose
got for Stillinger clusters in a Lennard-Jones system $z=2.1\pm 0.1$\cite
{Heyes1}. We found, averaging over all temperatures, $z_{Still}=2.05\pm 0.05$%
, $z_{Chem}=2.14\pm 0.05$, $z_{Phys}=2.30\pm 0.05$. As we can see, for
chemical clusters we have, within the statistical errors, a similar exponent
that for the Stillinger ones, whereas for physical clusters we observe that
the exponent value differs. We obtained the value of $z_{Phys}$ for physical
cluster with a longer life time, $\tau ^{*}=1.0,$ at $T^{*}=1.4$, for which
percolation density is $\rho _{p}^{*}=0.2\pm 0.01$ and $z_{Phys}=2.41\pm
0.04.$ This show that the critical exponent $z$ is very sensitive to the
life time selected for physical clusters.

Looking for structural information of the clusters we have calculated its
moments of inertia. Surprisingly, we get elongated clusters, no matter which
definition is used. Two of the principal moments of inertia (I$_{1}$ and I$%
_{2}$) are always similar between them and larger than the third one (I$_{3}$%
). This result was explained by Lar\'{i}a and Vericat\cite{Laria1} when
small clusters are under consideration. However, they got approximately
spherical big clusters for dipolar hard spheres. In Fig. 6 the ratio $%
(I_{1}+I_{2})/(2I_{3})$ is plotted as function of the cluster size for each
cluster definition. Physical clusters are the more spherical ones, with a
more near one ratio. Because big clusters are rare we average $%
(I_{1}+I_{2})/(2I_{3})$ for all clusters bigger than 20 particles size
obtaining $(I_{1}+I_{2})/(2I_{3})=2.52$. This is far away from the spherical
value.

The origin of the difference between our result and the result obtained by
Lar\'{i}a and Vericat lies in the fact that we use free boundary conditions
in the calculation of the inertia matrix\cite{Lee1}. Unphysical structures
are simulated when periodic boundary conditions are used as in the Lar\'{i}a
and Vericat work.

Recently, non-spherical clusters in nucleation studies have received special
interest. An experimental result shows that spherical particles aggregate in
non-spherical clusters at the liquid-solid transition\cite{Yau1}. However,
spherical clusters are expected for the gas-liquid transition\cite{Oxtoby1}.
Our simulation results show that at liquid densities non-spherical clusters
are the more frequent case.

We give here a very simple explanation, based on steric considerations, for
the existence of non-spherical clusters. For simplicity, suppose a
quasi-circular two-dimensional cluster [see Fig. 7(a)]. If we add a particle
in the accessible surface of the cluster a symmetry break down will occur.
Now, the accessible surface is greater on the right side of the cluster [see
Fig. 7(b)]. This increases the probability that a second particle added at
random lands in the right side of the cluster. So, the cluster tends to grow
in an elongated structure. Nevertheless, this process can not go on for
ever. If the cluster is very elongated, the lateral accessible surface [top
and bottom part in Fig. 7(c)] is very big. Then, the probability that a new
particle added at random lands in the lateral region dominate. The mean
shape of a cluster will balance the two competing effects. Suppose a cluster
with the shape of a spherical caped cylindre (cylindre length $L$ and caps
radius $R$) (see Fig. 8). The balance of the mentioned competing effects
means that the lateral surface of the cylinder has to be equal to the
surface of the caps. So, we chose $L=2R$. A solid revolution ellipsoid with
semiaxis $R$ and $2R$ has a moments of inertia ratio $%
(I_{1}+I_{2})/(2I_{3})=2.5$, which is quite similar to the value obtained
from our simulations for clusters bigger than 20 particles.

Another structural parameter is the radius of gyration $R_{g}.$ It is
defined for a cluster of size $s$ as\cite{Stauffer} 
\begin{equation}
R_{g}^{2}=\frac{1}{s}\sum_{i=1}^{s}|{\bf r}_{i}-{\bf r}_{cm}|^{2}  \label{n5}
\end{equation}
where ${\bf r}_{i}$ is the position of the ith particle in the cluster and $%
{\bf r}_{cm}$ is the position of the centre of mass of the cluster 
\begin{equation}
{\bf r}_{cm}=\frac{1}{s}\sum_{i=1}^{s}{\bf r}_{i}{}  \label{n6}
\end{equation}

In Fig. 9 we present the radius of gyration as a function of the size of the
cluster at the percolation density for each cluster definition. As usual $%
R_{g}$ follows the power law\cite{Stauffer} 
\begin{equation}
R_{g}\propto s^{1/D_{f}}  \label{n7}
\end{equation}
where $D_{f}$ is the fractal dimension of the clusters. In our simulation
that power law applies only for clusters bigger than 5 particles size. The
fractal dimension is the same for all clusters definitions. The chemical and
physical clusters are more compact than Stillinger clusters because they
have a smaller radius of gyration. In the calculation of $R_{g}$ we used
free boundary conditions as for all the previous structural quantities.

\section{Theory}

\subsection{Chemical clusters}

In this section we will summarize the main results of the theory that we
have developed elsewhere\cite{Pugnaloni1} to describe the clustering and
percolation for chemical clusters.

For a system of $N$ classical particles interacting via a pair potential $v(%
{\bf r}_{i},{\bf r}_{j})$ we define a density correlation function $\rho (%
{\bf r}_{1},{\bf r}_{2},{\bf p}_{1},{\bf p}_{2})$ which is $N(N-1)$ times
the probability density of finding two particles at phase space points $(%
{\bf r}_{1}$, ${\bf p}_{1})$ and $({\bf r}_{2}$, ${\bf p}_{2})$ respectively:

\begin{eqnarray}
\rho ({\bf r}_{1},{\bf r}_{2},{\bf p}_{1},{\bf p}_{2}) &=&\frac{N(N-1)}{%
h^{3N}N!Q_{N}(V,T)}  \label{n8} \\
&&\times \int \prod_{i=1}^{N}\exp [-\beta \frac{{\bf p}_{i}^{2}}{2m}%
]\prod_{i=1}^{N}\prod_{j>i}^{N}\exp [-\beta v({\bf r}_{i},{\bf r}_{j})]d{\bf %
r}^{N-2}d{\bf p}^{N-2}.  \nonumber
\end{eqnarray}
Here $h$ is the Planck constant and $Q_{N}(V,T)$ the canonical partition
function of the system. Then, in the same spirit of Hill and Coniglio et. al.%
\cite{Hill1,Coniglio1}, we separate $\exp [-\beta v({\bf r}_{i},{\bf r}%
_{j})] $ into connecting and blocking parts

\begin{equation}
\exp [-\beta v({\bf r}_{i},{\bf r}_{j})]=f^{\dagger }({\bf r}_{i},{\bf r}%
_{j},{\bf p}_{i},{\bf p}_{j})+f^{*}({\bf r}_{i},{\bf r}_{j},{\bf p}_{i},{\bf %
p}_{j})+1  \label{n9}
\end{equation}
Here $f^{\dagger }({\bf r}_{i},{\bf r}_{j},{\bf p}_{i},{\bf p}_{j})$
represent the basic probability density that two particles at configuration $%
({\bf r}_{i},{\bf r}_{j},{\bf p}_{i},{\bf p}_{j})$, are bonded. The
shorthand notation $f^{\gamma }({\bf r}_{i},{\bf r}_{j},{\bf p}_{i},{\bf p}%
_{j})\equiv f_{i,j}^{\gamma }$ ($\gamma =\dagger ,*$) will be sometimes used.

Substitution of Eq.(\ref{n9}) in Eq.(\ref{n8}) yields 
\begin{eqnarray}
\rho ({\bf r}_{1},{\bf r}_{2},{\bf p}_{1},{\bf p}_{2}) &=&\frac{N(N-1)}{%
h^{3N}N!Q_{N}(V,T)}\exp [-\beta v({\bf r}_{1},{\bf r}_{2})]  \label{n10} \\
&&\times \int \prod_{i=1}^{N}\exp [-\beta \frac{{\bf p}_{i}^{2}}{2m}]\sum
\{\prod f_{i,j}^{\dagger }f_{k,l}^{*}\}dr^{N-2}dp^{N-2},  \nonumber
\end{eqnarray}
where the sum is carried out over all possible arrangements of products of
functions $f_{i,j}^{\dagger }$ and $f_{k,l}^{*}$.

It should be noted that the functions $f_{i,j}^{\dagger }$ and $f_{i,j}^{*}$
can depend on momenta as well as on the positions of the two particles, but
the sum of $f_{i,j}^{\dagger }$ and $f_{i,j}^{*}$ must be momenta
independent in order that Eq.(\ref{n9}) be satisfied. Except by this last
condition, the functions $f_{i,j}^{\dagger }$ and $f_{i,j}^{*}$ are
otherwise arbitrary for thermodynamic purposes. Obviously we choose them in
such a way that the desired definition of bonded particles is achieved:

\begin{equation}
f^{\dagger }({\bf r}_{i},{\bf r}_{j},{\bf p}_{i},{\bf p}_{j})=%
{\exp [-\beta v({\bf r}_{i},{\bf r}_{j})] \atopwithdelims\{. 0}%
\begin{array}{c}
|{\bf r}_{i,j}(t)|<d\text{ \quad }\forall t\leq \tau \\ 
\qquad \quad otherwise
\end{array}
\label{n11}
\end{equation}

\begin{equation}
f^{*}({\bf r}_{i},{\bf r}_{j},{\bf p}_{i},{\bf p}_{j})=%
{-1 \atopwithdelims\{. \exp [-\beta v({\bf r}_{i},{\bf r}_{j})]-1}%
\begin{array}{c}
|{\bf r}_{i,j}(t)|<d\text{ \quad }\forall t\leq \tau \\ 
\qquad \quad otherwise
\end{array}
\label{n12}
\end{equation}
where ${\bf r}_{i,j}(t)$ is the relative position of particles $i$ and $j$
at time $t$. We see that, in fact, Eq.(\ref{n9}) is satisfied by Eqs. (\ref
{n11}) and (\ref{n12}). Explicitly, time is introduced here by taking the
set $\{{\bf r}^{N},{\bf p}^{N}\}$ as initial conditions in $t=0$ and solving
the equation of motion of the particles under their mutual interaction.

In order to exactly calculate ${\bf r}_{i,j}(t)$ for any $t$ we must solve a
many body problem. An approximation can be obtained by reducing it to a
two-body problem. This can be done by taking\cite{BenNaim1} the effective
force $[-\nabla v^{eff}({\bf r}_{i},{\bf r}_{j})]$ composed of the direct
force exerted by particle $j$ on $i$ $[-\nabla v({\bf r}_{i},{\bf r}_{j})]$
plus the conditional average force exerted by the remaining particles in the
system on the particle $i$. The last is an average over all configurations
of the $N-2$ particles keeping $i$ and $j$ fixed. Thus

\[
-\nabla v^{eff}({\bf r}_{i},{\bf r}_{j})=-\nabla v({\bf r}_{i},{\bf r}%
_{j})+\left\langle -\sum\limits_{k\neq i,j}^{N}\nabla v({\bf r}_{i},{\bf r}%
_{k})\right\rangle _{N-2}. 
\]
It can be shown\cite{BenNaim1} that this approximation corresponds to take $%
v^{eff}({\bf r}_{i},{\bf r}_{j})=-\ln [g({\bf r}_{i},{\bf r}_{j})]/\beta $.
This way ${\bf r}_{i,j}(t)$ is obtained in terms of just the initial values $%
{\bf r}_{i},{\bf r}_{j},{\bf p}_{i}$ and ${\bf p}_{j}$.

Equation (\ref{n11}) states that two particles $i$ and $j$ are bonded if
they are separated by a distance shorter than $d$ during a time interval
longer than $\tau $. This coincides with the definition introduced in
section II for chemical clusters.

Each term in the integrand of Eq.(\ref{n10}) can be represented as a diagram
consisting of two white $e_{1}$ and $e_{2}$-points, $N-2$ black $e_{i}$%
-points and some $f_{i,j}^{\dagger }$ and $f_{i,j}^{*}$ connections except
between the white points. Here we take $e_{i}\equiv \exp [-\beta \frac{{\bf p%
}_{i}^{2}}{2m}]$. White point are not integrated over whereas black points
are integrated over their positions and momenta. All the machinery normally
used to handle standard diagrams in classical liquid theory\cite{Hansen1}
can be now extended to treat these new kind of diagrams. By following the
Coniglio recipe to separate connecting and blocking parts in correlation
functions i.e. $g({\bf r}_{1},{\bf r}_{2})=g^{\dagger }({\bf r}_{1},{\bf r}%
_{2},{\bf p}_{1},{\bf p}_{2})+g^{*}({\bf r}_{1},{\bf r}_{2},{\bf p}_{1},{\bf %
p}_{2})$, we obtain an Ornstein-Zernike like integral equation for $%
g^{\dagger }({\bf r}_{1},{\bf r}_{2},{\bf p}_{1},{\bf p}_{2})$%
\begin{eqnarray}
g^{\dagger }({\bf r}_{1},{\bf r}_{2},{\bf p}_{1},{\bf p}_{2}) &=&c^{\dagger
}({\bf r}_{1},{\bf r}_{2},{\bf p}_{1},{\bf p}_{2})  \label{n13} \\
&&+\int \rho ({\bf r}_{3},{\bf p}_{3})c^{\dagger }({\bf r}_{1},{\bf r}_{3},%
{\bf p}_{1},{\bf p}_{3})g^{\dagger }({\bf r}_{3},{\bf r}_{2},{\bf p}_{3},%
{\bf p}_{2})d{\bf r}_{3}d{\bf p}_{3}.  \nonumber
\end{eqnarray}

Here $\rho ({\bf r}_{1},{\bf p}_{1})\rho ({\bf r}_{2},{\bf p}_{2})g^{\dagger
}({\bf r}_{1},{\bf r}_{2},{\bf p}_{1},{\bf p}_{2})$ is $N(N-1)$ times the
joint probability density of finding two particles at positions ${\bf r}_{1}$
and ${\bf r}_{2}$ with momenta ${\bf p}_{1}$ and ${\bf p}_{2}$,
respectively, and belonging to the same cluster, where the bonding criterion
is given by Eqs.(\ref{n11}) and (\ref{n12}); being 
\begin{equation}
\rho ({\bf r}_{1},{\bf p}_{1})=\frac{1}{N-1}\int \rho ({\bf r}_{1},{\bf r}%
_{2},{\bf p}_{1},{\bf p}_{2})d{\bf r}_{2}d{\bf p}_{2}.  \label{n14}
\end{equation}
The function $c^{\dagger }({\bf r}_{1},{\bf r}_{2},{\bf p}_{1},{\bf p}_{2})$
denotes the sum of all the non-nodal diagrams in the diagrammatic expansion
of $g^{\dagger }({\bf r}_{1},{\bf r}_{2},{\bf p}_{1},{\bf p}_{2}).$ We
remember that a nodal diagram contains at least a black point\thinspace
through which all paths between the two white points pass.

For homogeneous systems we have

\begin{eqnarray}
g^{\dagger }({\bf r}_{1},{\bf r}_{2},{\bf p}_{1},{\bf p}_{2}) &=&c^{\dagger
}({\bf r}_{1},{\bf r}_{2},{\bf p}_{1},{\bf p}_{2})+\frac{\rho }{(2\pi
mk_{B}T)^{3/2}}  \label{n15} \\
&&\times \int \exp [-\beta \frac{{\bf p}_{3}^{2}}{2m}]c^{\dagger }({\bf r}%
_{1},{\bf r}_{3},{\bf p}_{1},{\bf p}_{3})g^{\dagger }({\bf r}_{3},{\bf r}%
_{2},{\bf p}_{3},{\bf p}_{2})d{\bf r}_{3}d{\bf p}_{3}.  \nonumber
\end{eqnarray}

To get an integral equation from Eq.(\ref{n13}) is necessary a closure
relation between $g^{\dagger }({\bf r}_{1},{\bf r}_{2},{\bf p}_{1},{\bf p}%
_{2})$ and $c^{\dagger }({\bf r}_{1},{\bf r}_{2},{\bf p}_{1},{\bf p}_{2})$.
We take the Percus-Yevick approximation $g({\bf r}_{1},{\bf r}_{2})\exp
[\beta v({\bf r}_{1},{\bf r}_{2})]=1+N({\bf r}_{1},{\bf r}_{2}),$where the
function $N({\bf r}_{1},{\bf r}_{2})$ is the sum of the nodal diagrams in
the expansion of $g({\bf r}_{1},{\bf r}_{2})$. Separation into connecting
and blocking parts i.e. $g({\bf r}_{1},{\bf r}_{2})=g^{\dagger }({\bf r}_{1},%
{\bf r}_{2},{\bf p}_{1},{\bf p}_{2})+g^{*}({\bf r}_{1},{\bf r}_{2},{\bf p}%
_{1},{\bf p}_{2})$ and $N({\bf r}_{1},{\bf r}_{2})=N^{\dagger }({\bf r}_{1},%
{\bf r}_{2},{\bf p}_{1},{\bf p}_{2})+N^{*}({\bf r}_{1},{\bf r}_{2},{\bf p}%
_{1},{\bf p}_{2})$, yields 
\begin{eqnarray}
g^{\dagger }({\bf r}_{1},{\bf r}_{2},{\bf p}_{1},{\bf p}_{2}) &=&[f^{*}({\bf %
r}_{1},{\bf r}_{2},{\bf p}_{1},{\bf p}_{2})+1][g^{\dagger }({\bf r}_{1},{\bf %
r}_{2},{\bf p}_{1},{\bf p}_{2})-c^{\dagger }({\bf r}_{1},{\bf r}_{2},{\bf p}%
_{1},{\bf p}_{2})]  \label{n16} \\
&&+\exp [\beta v({\bf r}_{1},{\bf r}_{2})]g({\bf r}_{1},{\bf r}%
_{2})f^{\dagger }({\bf r}_{1},{\bf r}_{2},{\bf p}_{1},{\bf p}_{2}). 
\nonumber
\end{eqnarray}
Eq. (\ref{n13}) closed by Eq.(\ref{n16}) gives an integral equation for $%
g^{\dagger }({\bf r}_{1},{\bf r}_{2},{\bf p}_{1},{\bf p}_{2})$.

From the function $g^{\dagger }({\bf r}_{1},{\bf r}_{2},{\bf p}_{1},{\bf p}%
_{2})$ we define the pair correlation function for chemical clusters

\begin{equation}
g_{Chem}({\bf r}_{1},{\bf r}_{2})=\int \rho ({\bf r}_{1},{\bf p}_{1})\rho (%
{\bf r}_{2},{\bf p}_{2})g^{\dagger }({\bf r}_{1},{\bf r}_{2},{\bf p}_{1},%
{\bf p}_{2})d{\bf p}_{1}d{\bf p}_{2}.  \label{n17}
\end{equation}
It is the joint probability density of finding two particles that belong to
the same chemical cluster at positions ${\bf r}_{1}$ and ${\bf r}_{2}$,
respectively.

Then, the mean cluster size $S_{chem}$ is given by Eq.(\ref{n1}) and the
percolation density is calculated from Eq.(\ref{n2}) with $\gamma =Chem$:

\[
S_{Chem}=1+\frac{1}{(N-1)}\int g_{Chem}({\bf r}_{1},{\bf r}_{2})d{\bf r}_{1}d%
{\bf r}_{2}, 
\]
\[
\lim\limits_{\rho \rightarrow \rho _{c}^{-}}S_{Chem}(\rho )=\infty . 
\]

\subsection{Weak criterion for physical clusters}

In order to treat the physical clusters within a theoretical framework we
will restrict ourself to a weak version of the definition introduced in
section III. The weak criterion is established as follows

\begin{center}
{\bf Two particles of the system belong to the same physical cluster at the
time }$t${\bf \ if at that time exists a path of directly connected
particles which link them, and all particles forming the path, including the
two particles under study, were connected in time }$t-\tau .$
\end{center}

Connected and directly connected are understood in the Stillinger sense. The
difference with the strong criterion established in Section III is that in
the weak version we require that the particles be connected just at the
extremes of the time interval and not during the whole interval. The idea
for the future is to use the results obtained from this definition within a
path-integral like formalism in order to describe clusters according to the
strong criterion.

Consider the function $H({\bf r}_{1},{\bf r}_{2})$ introduced by Xu and
Stell in their ``percolation in probability'' theory\cite{Xu1}. This
function measures the basic conditional probability for two particles to be
directly connected if they are at positions ${\bf r}_{1}$ and ${\bf r}_{2}$,
respectively. The role of this function within Hill formalism is that of an
auxiliary function in the Boltzmann factor separation into bonded and
unbounded terms: 
\begin{eqnarray}
e_{i,j}^{\dagger } &=&H({\bf r}_{i},{\bf r}_{j})e({\bf r}_{i},{\bf r}_{j})
\label{n18} \\
e_{i,j}^{*} &=&\left[ 1-H({\bf r}_{i},{\bf r}_{j})\right] e({\bf r}_{i},{\bf %
r}_{j}).  \nonumber
\end{eqnarray}

Each choice of $H({\bf r}_{i},{\bf r}_{j})$ corresponds to a different bond
definition. For example, for the Stillinger's definition of directly
connected particles we have

\[
H({\bf r}_{1},{\bf r}_{2})=H(r_{1,2})=\left\{ 
\begin{array}{ll}
1 & r_{1,2}\leq d \\ 
0 & r_{1,2}>d
\end{array}
\right. . 
\]

To define bonded particles according with the weak criterion of physical
clusters, we use: 
\begin{equation}
H({\bf r}_{1},{\bf r}_{2})=H(r_{1,2};\tau )=\left\{ 
\begin{array}{ll}
\frac{Nh^{\dagger }({\bf r}_{1,2};\tau )}{\rho g({\bf r}_{1,2})} & 
r_{1,2}\leq d \\ 
0 & r_{1,2}>d
\end{array}
\right. ,  \label{n19}
\end{equation}
where $h^{\dagger }({\bf r}_{1,2};\tau )$ is the probability density of
finding two particles connected at time $t-\tau $ and in relative position $%
{\bf r}_{1,2}$ at time $t$. Clearly, this selection of $H(r_{1,2};\tau )$
considers as bonded any two particles which are separated by a distance
shorter than $d$ and have been connected (directly or not) at time $t-\tau $%
. We explicitly include $\tau $ in the functions to stress the life time
dependence.

Once the direct bond is defined, a connectivity Ornstein-Zernike relation
between the pair correlation function $g_{Phys}(r_{1,2};\tau )$ and the
direct correlation function for physical clusters $c_{Phys}(r_{1,2};\tau )$
can be obtained in the standard form\cite{Hansen1,Coniglio2}:

\begin{equation}
g_{Phys}(r_{1,2};\tau )=c_{Phys}(r_{1,2};\tau )+\rho \int
c_{Phys}(r_{1,3};\tau )g_{Phys}(r_{3,2};\tau )d{\bf r}_{3}.  \label{n20}
\end{equation}
The function $g_{Phys}(r_{1,2};\tau )$ is $N(N-1)$ times the probability
density of finding two particles in positions ${\bf r}_{1}$ and ${\bf r}_{2}$
and belonging to the same physical cluster.

In order to solve equation (\ref{n20}) a closure relation is needed. For $%
r_{1,2}\leq d$, we see that $g_{Phys}(r_{1,2};\tau )$ is exactly equal to $%
\frac{N}{\rho }h^{\dagger }({\bf r}_{1,2};\tau )$, because the conditional
probability density that two particles belong to the same physical cluster
under the assumption they were connected at $t-\tau $ is equal to one if
they are separated by a distance shorter than $d$. For $r_{1,2}>d$ we can
use the Percus-Yevick connectivity closure\cite{Coniglio2}.

The calculation of $h^{\dagger }(r_{1,2};\tau )$ remains to be discussed.
Consider the temporal correlation function introduced by Oppenheim and Bloom
in the study of nuclear spin relaxation $G({\bf r}_{1,2}^{\prime },{\bf r}%
_{1,2},t)$\cite{Oppenheim1}. This function is the probability density of
finding two particles in relative position ${\bf r}_{1,2}^{\prime }=$ ${\bf r%
}_{2}^{\prime }-$ ${\bf r}_{1}^{\prime }$ at an initial time and in relative
position ${\bf r}_{1,2}=$ ${\bf r}_{2}-$ ${\bf r}_{1}$at time $t$ later. It
is given by 
\begin{eqnarray}
G({\bf r}_{1,2}^{\prime },{\bf r}_{1,2},t) &=&\frac{1}{N(N-1)}  \label{n21}
\\
&&\times \left\langle \sum\limits_{i}^{N}\sum\limits_{j\neq i}^{N}\delta
\left[ {\bf r}_{1,2}^{\prime }-\left( {\bf r}_{i}(0)-{\bf r}_{j}(0)\right)
\right] \delta \left[ {\bf r}_{1,2}-\left( {\bf r}_{i}(t)-{\bf r}%
_{j}(t)\right) \right] \right\rangle ,  \nonumber
\end{eqnarray}
where $\delta (x)$ is Dirac's delta function and $\left\langle
.\right\rangle $ mean an average over all initial configurations in the
canonical ensemble. In the case of an ideal gas, i.e. non-interacting
particles of mass $m$, $G({\bf r}_{1,2}^{\prime },{\bf r}_{1,2},t)$ is
exactly given by

\begin{equation}
G({\bf r}_{1,2},{\bf r^{^{\prime }}}_{1,2},\tau )=\frac{\rho }{N}\left( 
\frac{\beta m}{4\pi \tau ^{2}}\right) ^{3/2}\exp \left[ -\frac{\beta m({\bf r%
}_{1,2}-{\bf r^{^{\prime }}}_{1,2})^{2}}{4\tau ^{2}}\right] .  \label{n22}
\end{equation}
For systems with non-null interaction some approximations for $G({\bf r}%
_{1,2}^{\prime },{\bf r}_{1,2},t)$ have been reported\cite
{Oppenheim1,Bloom1,Balucani1}.

Now we define 
\begin{equation}
h^{\dagger }({\bf r}_{1,2};\tau )=\int \frac{g_{Still}(r_{1,2}^{\prime })}{%
g(r_{1,2}^{\prime })}G({\bf r}_{1,2}^{\prime },{\bf r}_{1,2},\tau )d{\bf r}%
_{1,2}^{\prime },  \label{n23}
\end{equation}
where $g_{Still}(r_{1,2}^{\prime })$ is the pair correlation function for
Stillinger clusters and $g(r_{1,2}^{\prime })$ is the ordinary pair
correlation function. The function $h^{\dagger }({\bf r}_{1,2};\tau )$ is
proportional to the probability density of finding two particles connected
at time $t-\tau $ and in relative position ${\bf r}_{1,2}$ at time $t$. Note
that in the final time $t$ the particles can be connected or not, but they
must be in relative position ${\bf r}_{1,2}$, whereas in the initial time
they must be connected whatever the relative position is.

We have applied the previous formalism to an ideal gas. In this case $%
g(r_{1,2})\equiv 1$ and $G({\bf r}_{1,2},{\bf r^{^{\prime }}}_{1,2},\tau )$
is given by Eq.(\ref{n22}).

The correlation function for Stillinger clusters $g_{Still}(r_{1,2})$ was
obtained, for this system, by Chiew and Glandt in the Percus-Yevick
approximation [$c^{\dagger }(r_{1,2})=0$ for $r_{1,2}>d$]\cite{Chiew1}. We
used this solution to obtain $h^{\dagger }(r_{1,2};\tau )$ from equation (%
\ref{n23}). As we have mentioned, $g_{Phys}(r_{1,2};\tau )$ is known exactly
for$\;r_{1,2}\leq d$. To close equation (\ref{n20}) we take the
approximation: $c_{Phys}(r_{1,2};\tau )=0$ for$\;r_{1,2}>d$. By using the
Baxter's factorization method\cite{Baxter1} and the Perram's algorithm\cite
{Perram1} we have solved the integral equation for $g_{Phys}(r_{1,2};\tau )$
in$\;r_{1,2}>d$.

In order to analyze the approximations we have introduced in our theoretical
approach, we have also simulated by molecular dynamics the non-interacting
system and have obtained the pair correlation function for physical clusters
defined in the weak sense. The simulation algorithm to identify the weak
physical clusters is slightly different from that considered in Subsection
IV-B for the strong criterion. It is carried out as follow:

{\bf 1)} Identify a Stillinger cluster at $t-\tau $.

{\bf 2)} Move the particles of this cluster to the new positions in $t$
without using periodic boundary conditions. In a ideal gas periodic boundary
conditions are unphysical.

{\bf 3)} Identify the Stillinger clusters in the new configuration for the
particles selected in {\bf 1. }These clusters are the physical clusters.

{\bf 4)} Repeat from {\bf 1} until cover all initial Stillinger clusters.

Fig. 10 shows the function $g_{Phys}(r_{1,2};\tau )$, obtained from our
theory and simulation for the weak physical clusters at reduced density $%
\rho ^{*}=\rho d^{3}=0.2$ and various reduced life times $\tau $$^{*}=\tau
/(d\sqrt{\beta m})$. When $\tau ^{*}\rightarrow 0$ we recover the Stillinger
clusters $g_{Phys}(r_{1,2};\tau =0)=g_{Still}(r_{1,2})$.

In Fig. 11 we present the mean cluster size for the physical clusters as a
function of $\tau ^{*}$ for various values of $\rho ^{*}$ obtained from Eq.(%
\ref{n1}) with $\gamma =Phys$: 
\[
S_{Phys}=1+\frac{1}{(N-1)}\int g_{Phys}({\bf r}_{1},{\bf r}_{2})d{\bf r}_{1}d%
{\bf r}_{2}. 
\]
At low densities, $\rho ^{*}<0.3$, the theory fits very well the simulation
results. For higher densities the approximations used for $%
g_{Still}(r_{1,2}) $ and $c_{Phys}(r_{1,2};\tau )$ for$\;r_{1,2}>d$ lead to
poor results.

\section{Comments}

\subsection{Summary and further developments}

In this work we have introduced two criteria for the cluster identification
in continuum systems. Both criteria are based in the dynamics of the system.
The first one uses a residence time in the definition of the bond between
pairs of particles, the second one uses a life time in the definition of an
aggregate. Due to the qualitative features of the clusters yielded by the
criteria we call them chemical and physical clusters, respectively.

The main results that show the differences between chemical, physical and
the traditional Stillinger clusters are collected below.

{\bf Physical clusters:}

{\bf a)} The percolation curve presents a slope similar to the corresponding
curve for Stillinger clusters. However, this curve locates at higher
densities than for the Stillinger case.

{\bf b)} The clusters are more spherical than the Stillinger ones for sizes
smaller than 20 particles. For clusters bigger than 20 particles size,
physical and Stillinger clusters have the same shape and are always
elongated.

{\bf c)} Big clusters are rare in comparison with Stillinger clusters. At
percolation density a power law is followed by the cluster size
distribution, but with a strongly $\tau $-dependent critical exponent.

{\bf d)} The fractal dimension of the clusters is the same that for
Stillinger clusters but the physical ones are more compact.

{\bf Chemical clusters:}

{\bf a)} The percolation curve present a small slope as compared with the
corresponding Stillinger curve. If the same value of $\tau $ is used for
physical and chemical clusters, then the percolation curve appears at higher
densities in the chemical case.

{\bf b)} The sphericity of the chemical clusters is in between the physical
and the Stillinger ones.

{\bf c)} At the percolation density a power law is followed by the cluster
size distribution with an almost $\tau $-independent critical exponent.

{\bf d)} The fractal dimension of the clusters is the same that for
Stillinger clusters but the chemical ones are as compact as the physical
clusters.

The definitions used in this work restricts us to use molecular dynamics
simulation, because the trajectory of the particles must be known. To
determine which particle belong to a given clusters at a given time $t$
simulations need to be run between $t-\tau $ and $t$ in very short time
steps. This takes a large computing time. Thus, approximate theories are
very much desirable to be developed.

The solution of Eq. (\ref{n13}) for chemical clusters is a very difficult
task. Tensors of rank 9 must be used to storage correlation functions and a
double convolution has to be solved. Recently, F. Lado\cite{Lado1} has
developed a method to calculate the pair correlations for a polarizable
Lennard-Jones system which is based in expanding the relevant functions in
orthogonal polynomials. The integral equation he has to solve is similar to
Eq. (\ref{n13}) but with ${\bf p}_{1}$ and ${\bf p}_{2}$ representing the
induced dipolar moments of the particles 1 and 2 respectively. We are
currently working in the application of Lado's code to the chemical
clustering problem for Lennard-Jones particles.

In the theory for physical clusters the main difficulty is to have a good
approximation for $G({\bf r}_{1,2}^{\prime },{\bf r}_{1,2},t)$ and $%
g_{Still}(r_{1,2})$. In one hand, the present theories to calculate the
correlation function for Stillinger clusters do not provide it for densities
higher than the percolation density. Because percolation densities are
higher for physical than for Stillinger clusters the theory developed here
is unable to give results for densities close to the ``physical''
percolation density. In other hand, approximations for $G({\bf r}%
_{1,2}^{\prime },{\bf r}_{1,2},t)$ in interacting systems are available just
for the limiting cases $t\approx 0$ and $t\approx \infty $ \cite
{Oppenheim1,Bloom1,Balucani1,Balucani2}.

Because the theory developed for physical clusters correspond to a weak
criterion, an improved theory to describe the physical clusters introduced
in section III is desirable. We hope that using a path integral approach in
the interval $\left[ t-\tau ,t\right] ,$ where the weak criterion is used in
each step of the path, the ``strong'' criterion could be reached.

\subsection{Potential applications}

Water in oil microemulsions are formed by mixing water, oil and a
surfactant. Surfactant configuration favors the formation of inverted
micelles with water inside and oil external to spherical structures\cite
{Chen1}. When the micellar concentration is low, the system is a poor
electrical conductor. Nevertheless, when the concentration of micelles
increases beyond a critical value a conducting behavior is present. This
conductor-insulator transition has been explained by a continuum percolation
model\cite{Cametti1,Chen1}. When an aggregate of micelles spans the system,
charge carriers hop from micelle to micelle in the aggregate and the system
becomes a conductor. The process of passing a charged carrier from a given
micelle to another one to which the former is directly connected requires
some time. One can think of at least two ways of modeling this feature. In
one of them the real pair attraction between micelles is exaggerated using,
for example, sticky potentials to mimic the real intermicellar bonds. In the
other one, we consider a pair potential closer to the real one and require
the two micelles be connected during a time longer than $\tau $. This last
point of view agrees with the idea of a chemical cluster.

Most of the work on conductor/transition in microemulsions has followed the
first approach. Thus, the percolation curve for Stillinger clusters in a
sticky hard spheres system was shown to fit the experimental
conductor-insulator curve for
water/decane/sodium-bis(2-ethylhexil)sulfosuccinate\cite{Chen1}. The curve
has a peculiar behavior: it presents a small slope as compared with any
other system\cite{Chiew1,Seaton1}. This behavior coincides with that
presented by the chemical clusters in a Lennard-Jones system as we expect
from the considerations above: the sticky spheres can be defined as bonded
by just requiring a geometrical restriction in a given instant. As the
interaction potential is infinitely deep at contact no residence time is
necessary to impose in order to ensure a real bond. However, a continuum
pair potential is a more realistic interaction for micelles and, in that
case, the Stillinger criterion is not able to describe the
conductor-insulator transition for water in oil microemulsions. In
conclusion, the hopping time, e.g. the minimum time two micelles have to be
close in order for a charge carrier to hop between them, seems to be
necessary if realistic intermicellar pair potentials are going to be
considered. The chemical clusters provide an accurate model to describe the
conductor-insulator transition in this case. The values of $d$ and $\tau $,
for a chemical bond in Section II, can be used for water-in-oil
microemulsions as the maximum distance and minimum time required to allow
charge carriers to hop between two given micelles.

As we mentioned, the concept of cluster is of fundamental importance in
nucleation studies on saturated vapors. These theories concern the growth of
liquid-density-aggregates in a saturated vapor, which are the seed of a
liquid phase. A comparative study of different clustering criteria was done
recently by Senger et. al.\cite{Senger1}. The most frequently used criterion
for clusters identification in nucleation studies is the Stillinger
criterion and some improved versions of it. Nevertheless, this criterion can
wrongly consider as part of a cluster some gas particles which are close to
the cluster for a short time. In this way the cluster density will be lower
than the density of the cluster obtained by considering that those particles
do not belong to it. Also, the Stillinger criterion consider two physical
clusters, which are close together for a very short time, as a single cluster%
\cite{Abraham1}. And a single cluster formed of two real clusters joined by
only a point during just a very short period of time, will have a density of
particles that is lower than the corresponding density inside each of the
two separated clusters.

Several modifications have been used to avoid this unphysical cluster
counting\cite{Senger1,Rein1}. However, the inclusion of dynamical
information in the identification of clusters has been proposed as essential
in selecting long-lived clusters in nucleation studies\cite{Senger2}.
Although we have studied systems which are in equilibrium, it is possible to
use the previous clusters definitions in computer simulations of
non-equilibrium systems also. We hope the physical clusters introduced in
this work could contribute to a more appropriate identification of the
relevant clusters in nucleation because their existence depends on a life
time restriction.

{\bf ACKNOWLEDGMENTS}

We wish to thank Dr. O.H. Scalise for critical reading of the maniscript.
Support of this work by Universidad Nacional de La Plata (Grant I055),
Consejo Nacional de Investigaciones Cient\'{i}ficas y T\'{e}cnicas (PIP
4690) and Agencia Nacional de Promoci\'{o}n Cient\'{i}fica y Tecnol\'{o}gica
(PICT 03-04517) of Argentina is very much appreciated. \bigskip L.A.P. is a
fellow and F.V is a member of CONICET.

\begin{center}
{\bf Figures Captions}
\end{center}

{\bf Figure 1:} Physical cluster identification. a) Position of the
particles at $t-\tau $. The two Stillinger clusters are painted in different
ways. b) Position of the particles at $t$. The two new Stillinger clusters
are painted in different ways. c) As in (b) but the four physical clusters
are painted in different ways. d) As in (b) but the five chemical clusters
are shown.

{\bf Figure 2:} Correlation function for the clusters at $T^{*}=1.4$ and $%
\rho ^{*}=0.155$. Circles, triangles and diamonds correspond to Stillinger,
chemical and physical clusters respectively. The doted line is the
(ordinary) radial distribution function. The selected density is the
percolation density for Stillinger clusters.

{\bf Figure 3:} Percolation curve for the Lennard-Jones fluid. Circles,
triangles and diamonds correspond to Stillinger, chemical and physical
clusters respectively. Solid squares correspond to the coexistence curve$%
^{39}$

{\bf Figure 4:} Log-Log plot of the cluster size distribution. Solid, doted
and dashed line correspond to Stillinger, chemical and physical clusters
respectively. $T^{*}=1.4$ and $\rho ^{*}=0.155$ in all cases.

{\bf Figure 5:} Log-Log plot of the cluster size distribution. Circles,
triangles and diamonds correspond to Stillinger, chemical and physical
clusters respectively. In each case the percolation density for $T^{*}=1.4$
was chosen. The $z$ values were obtained by linear regression.

{\bf Figure 6:} Ratio between the largest (I$_{1}$ and I$_{2}$) and smaller
(I$_{3}$) moments of inertia as function of the cluster size. Circles,
triangles and diamonds correspond to Stillinger, chemical and physical
clusters respectively. The horizontal solid line is the averaged value over
all clusters larger than 20 particles size.

{\bf Figure 7:} Two dimensional aggregates..

{\bf Figure 8:} Average shape of a big cluster.

{\bf Figure 9:} Log-Log plot of the radius of gyration as a function of the
cluster size at $T^{*}=1.4$ and percolation density. Circles, triangles and
diamonds correspond to Stillinger, chemical and physical clusters
respectively. The fractal dimension was obtained by linear regression.

{\bf Figure 10:} $g_{Phys}(r_{1,2};\tau )$ vs. $r_{1,2}$ for $\rho ^{*}=0.2$%
, theory (solid line) and simulation (symbols). Diamonds, circles, triangles
and squares correspond the $\tau ^{*}=0,0.1,0.316$ and $1.0$ respectively.
Note the scale change at $r_{1,2}/d=1.$

{\bf Figure 11:} Inverse of the mean cluster size for physical clusters as a
function of $\tau ^{*}$, theory (solid line) and simulation (symbols).
Squares, circles and triangles correspond to $\rho ^{*}=0.2,0.3$ and $0.5$
respectively.


\vspace{0in}

\begin{references}
\bibitem{Bijl1}  A. Bijl, Doctoral dissertation, Leiden (1938).

\bibitem{Band1}  W. Band, J. Chem. Phys. {\bf 7}, 324 (1939).

\bibitem{Frenkel1}  J. Frenkel, J. Chem. Phys. {\bf 7}, 538 (1939).

\bibitem{Mayer1}  J.E. Mayer, J. Chem. Phys. {\bf 5}, 67 (1937).

\bibitem{Hill1}  T.L. Hill, J. Chem. Phys. {\bf 23}, 617 (1955).

\bibitem{Stillinger1}  F. H. Stillinger, J. Chem. Phys. {\bf 38}, 1486
(1963).

\bibitem{Sciortino1}  F. Sciortino, P.H. Poole, H.E. Stanley and S. Havlin,
Phys. Rev. Lett. {\bf 64}, 1686 (1990).

\bibitem{Senger1}  B. Senger, P. Schaaf, D. S. Corti, R. Bowles, J. C.
Voegel and H. Reiss, J. Chem. Phys. {\bf 110}, 6421 (1999).

\bibitem{Schenter1}  G. K. Schenter, S. M. Kathmann and B. C. Garrett, Phys.
Rev. Lett. {\bf 82}, 3484 (1999).

\bibitem{Kusaka1}  I. Kusaka, Z. G. Wang and J. H. Seinfeld, J. Chem. Phys. 
{\bf 108}, 3416 (1998).

\bibitem{Cametti1}  C. Cametti, P, Codastefano, P. Tartaglia, S. H. Chen and
J. Rouch, Phys. Rev. A {\bf 45}, R5358 (1992).\ 

\bibitem{Mayer2}  J.E. Mayer and E. Montroll, J. Chem. Phys. {\bf 9}, 2
(1941).

\bibitem{Coniglio1}  A. Coniglio, U. De Angelis, A. Forlani and G. Lauro, J.
Phys. A: Math. Gen. {\bf 10}, 219 (1977).

\bibitem{Stell1}  G. Stell, J. Phys. A: Math. Gen. {\bf 17}, L885 (1984).

\bibitem{Chiew1}  Y.C. Chiew and E.D. Glandt, J. Phys. A: Math. Gen. {\bf 16}%
, 2599 (1983).

\bibitem{Chiew2}  Y.C. Chiew, G. Stell and E.D. Glandt, J. Chem. Phys. {\bf %
83}, 761 (1985).

\bibitem{Gawlinski1}  E.T. Gawlinski and E.H. Stanley, J. Phys. A: Math.
Gen. {\bf 14}, 1291 (1981).

\bibitem{Sevick1}  E.M. Sevick, P.a. monson and J.M. Ottino, J. Chem. Phys. 
{\bf 88}, 1198 (1988).

\bibitem{DeSimone1}  T. DeSimone, S. Demoulini and R.M. Stratt, J. Chem.
Phys. {\bf 85}, 391 (1986).

\bibitem{Lee1}  S.B. Lee and S. Torquato, J. Chem. Phys. {\bf 89}, 6427
(1988).

\bibitem{Seaton1}  N.A. Seaton and E.D. Glandt, J. Chem. Phys. {\bf 86},
4668 (1987).

\bibitem{Netemeyer1}  S.C. Netemeyer and E.D. Glandt, J. Chem. Phys. {\bf 85}%
, 6054 (1986).

\bibitem{Chiew3}  Y.C. Chiew and Y.H. Wang, J. Chem. Phys. {\bf 89}, 6385
(1988).

\bibitem{Xu1}  J. Xu and G. Stell, J. Chem. Phys. {\bf 89}, 1101 (1988).

\bibitem{Bresme1}  F. Bresme and J. L. F. Abascal, J. Chem. Phys. {\bf 99},
9037 (1993).

\bibitem{Heyes1}  D. M. Heyes and J. R. Melrose, Mol. Phys. {\bf 66}, 1057
(1989).

\bibitem{Vericat1}  F. Vericat, J. Phys.: Condensed Matter, {\bf 1}, 5202
(1989); {\bf 2}, 3697 (1990).

\bibitem{Laria1}  D. Lar\'{i}a and F. Vericat, Phys. Rev. A {\bf 43}, 1932
(1991).

\bibitem{Carlevaro1}  C.M. Carlevaro, C. Stoico and F. Vericat, J. Phys.:
Condensed Matter, {\bf 8}, 1857 (1996).

\bibitem{Pugnaloni1}  L. A. Pugnaloni and F. Vericat, Phys. Rev. E {\bf 61},
R6067 (2000).

\bibitem{Sciortino2}  F. Sciortino and S. L. Forlini, J. Chem. Phys. {\bf 90}%
, 2786 (1989).

\bibitem{Abraham1}  F. F. Abraham and J. A. Barker, J. Chem. Phys. {\bf 63},
2266 (1975).

\bibitem{Luzar1}  A. Luzar and D. Chandler, Phys. Rev. Lett. {\bf 76}, 928
(1996).

\bibitem{Harrington1}  S. Harrington, R. Zhang, P.H. Poole, F. Sciortino and
H.E. Stanley, Rev. Lett. {\bf 78}, 2409 (1997).

\bibitem{Starr1}  F.W. Starr, J.K. Nielsen and H.E. Stanley, Phys. Rev.
Lett. {\bf 82}, 2294 (1999).

\bibitem{Nota1}  Chemical clusters are not useful if we want to study
aggregates where no bonded pairs exist.

\bibitem{Senger2}  B. Senger, P. Schaaf, D. S. Corti, R. Bowles, D. Pointu,
J. C. Voegel and H. Reiss, J. Chem. Phys. {\bf 110}, 6438 (1999).

\bibitem{Allen1}  M. P. Allen and D. J. Tildesley, {\it Computer Simulation
of Liquids} (Clarendon Press, Oxford, 1987).

\bibitem{Stoddard1}  S. D. Stoddard, J. Comp. Phys. {\bf 27}, 291 (1978).

\bibitem{Hoshen1}  J. Hoshen and R. Kopelman, Phys. Rev. B {\bf 28}, 5323
(1983).

\bibitem{Panagio1}  A. Z. Panagiotopoulos, Mol. Phys. {\bf 61}, 813 (1987).

\bibitem{Stauffer}  D. Stauffer, {\it Introduction to percolation theory}
(London: Taylor \& Francis, 1985).

\bibitem{Yau1}  S. T. Yau and P. G. Vekilov, Nature {\bf 406}, 494 (2000).

\bibitem{Oxtoby1}  D. W. Oxtoby, Nature {\bf 406}, 464 (2000).

\bibitem{BenNaim1}  A. Ben Naim, {\it Water and aqueous solutions }(New
York: Plenum Press, 1974)

\bibitem{Hansen1}  J.P. Hansen and I.R. McDonald, {\it Theory of Simple
Liquids} (Academic Press, London, 1976).

\bibitem{Coniglio2}  A. Coniglio, U. De Angelis and A. Forlani, J. Phys. A:
Math. Gen. {\bf 10}, 1123 (1977).

\bibitem{Oppenheim1}  I. Oppenheim and M. Bloom, Can. J. Phys., {\bf 39},
845 (1961).

\bibitem{Bloom1}  M. Bloom\ and I. Oppenheim, Adv. Chem. Phys. {\bf 12}, 549
(1967).

\bibitem{Balucani1}  U. Balucani and R. Vallauri, Phys. Lett. A, {\bf 76},
223 (1980).

\bibitem{Balucani2}  U. Balucani and R. Vallauri, Physica, {\bf 102A}, 70
(1980).

\bibitem{Baxter1}  R.J. Baxter, Aust. J. Phys., {\bf 21}, 563 (1968); J.
Chem. Phys. {\bf 52}, 4559 (1970).

\bibitem{Perram1}  J.W. Perram, Mol. Phys. {\bf 30}, 1505 (1975).

\bibitem{Lado1}  F. Lado, Phys. Rev. E {\bf 55}, 426 (1997).

\bibitem{Chen1}  S. H. Chen, J. Rouch, F. Sciortino and P. Tartaglia, J.
Phys.: Condens. Matter {\bf 6}, 10855 (1994).\ 

\bibitem{Rein1}  P. Rein ten Wolde and D. Frenkel, J. Chem. Phys. {\bf 109},
9901 (1998).
\end{references}
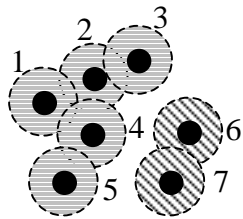
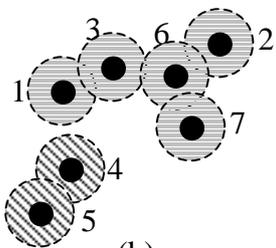
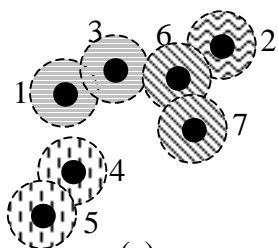
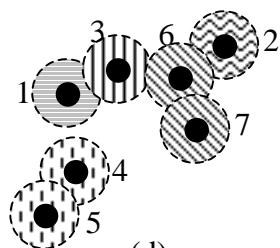
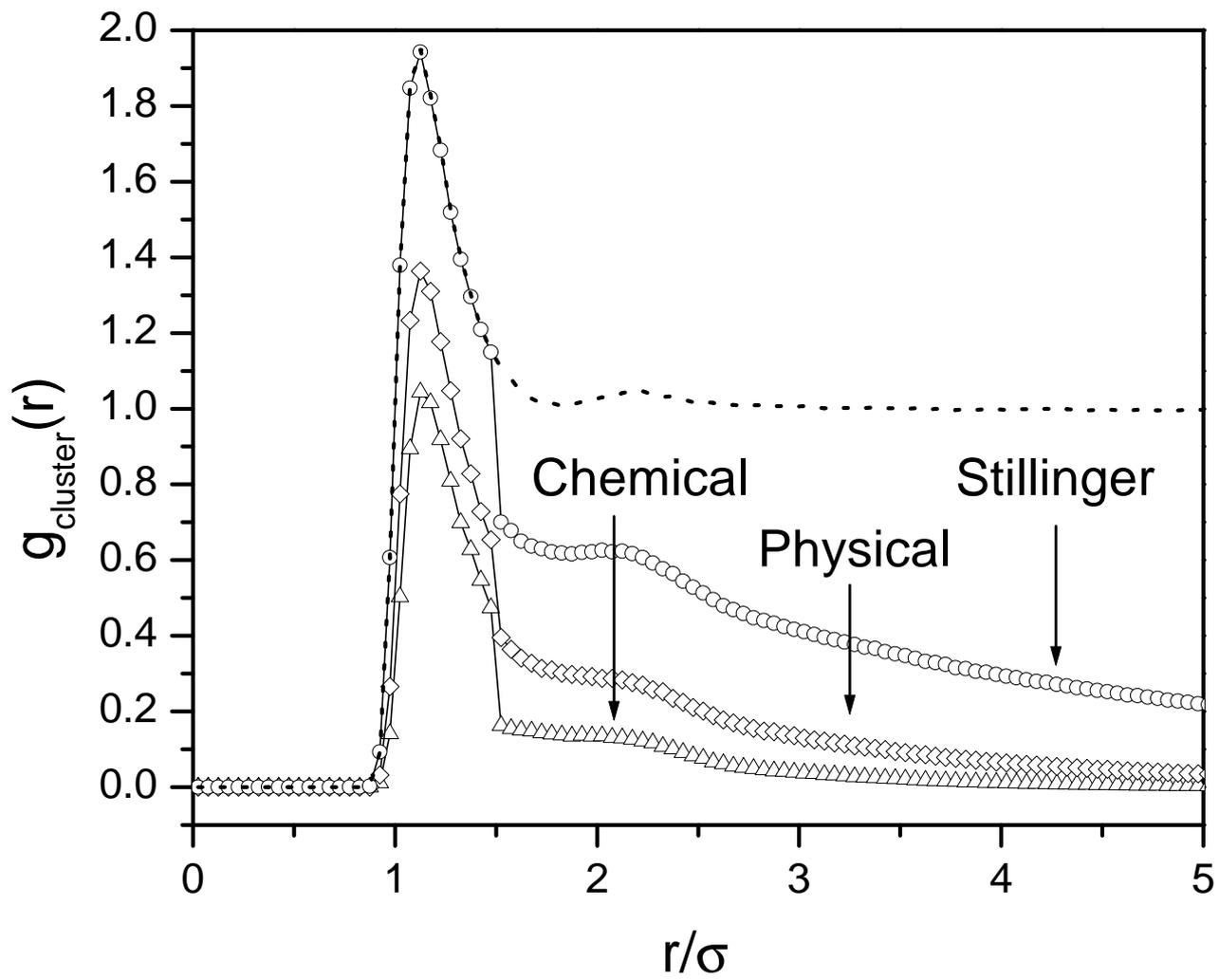
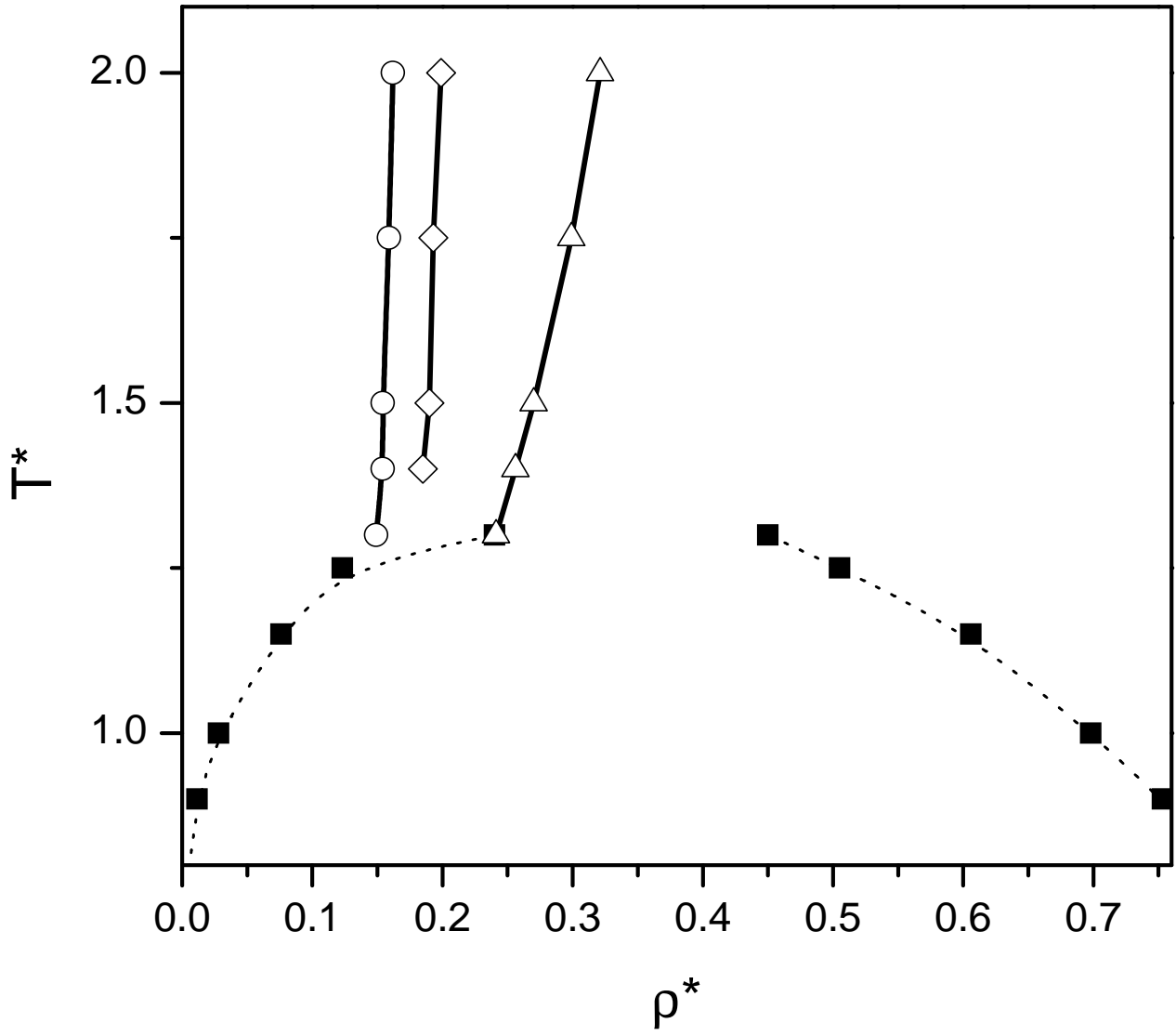
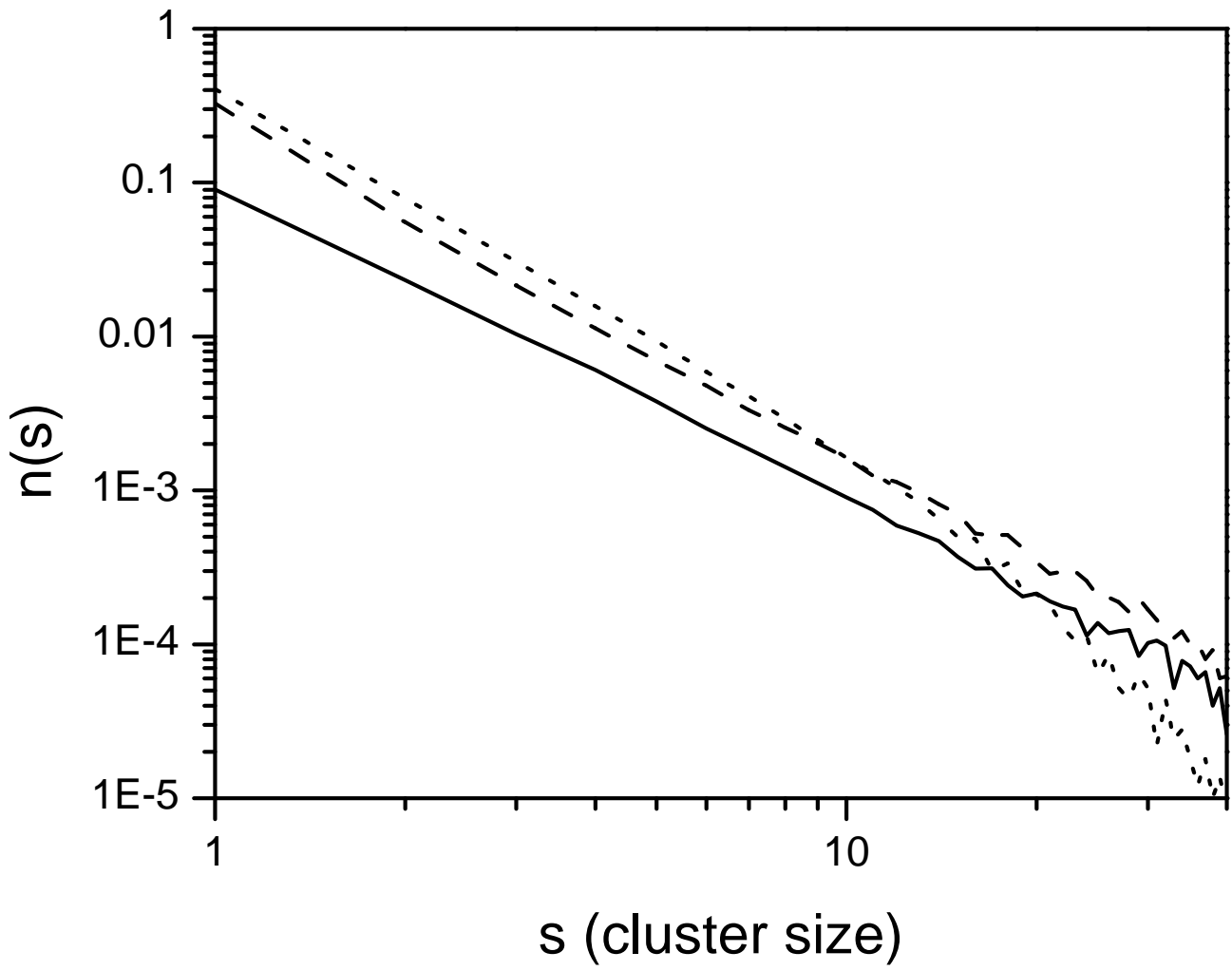
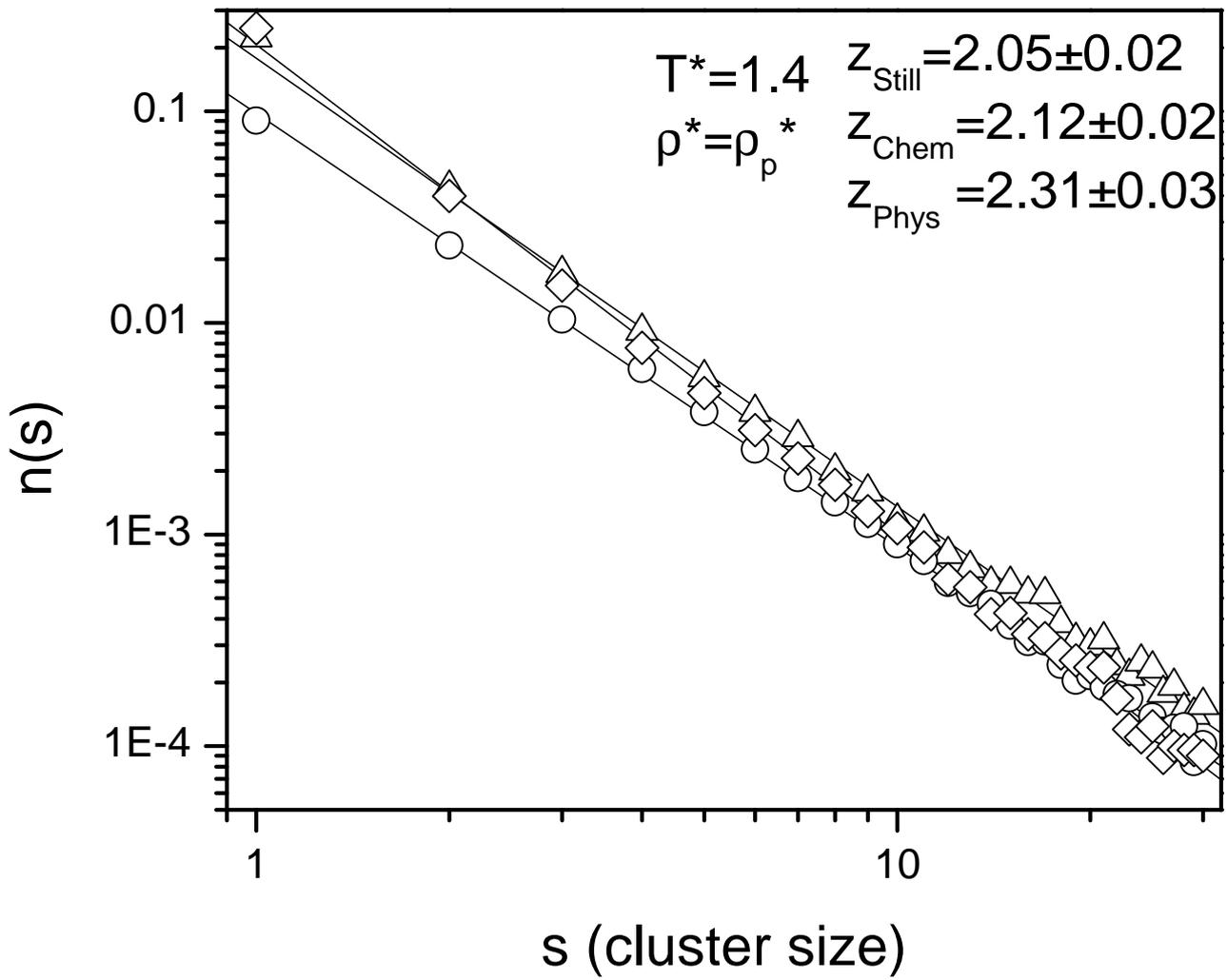
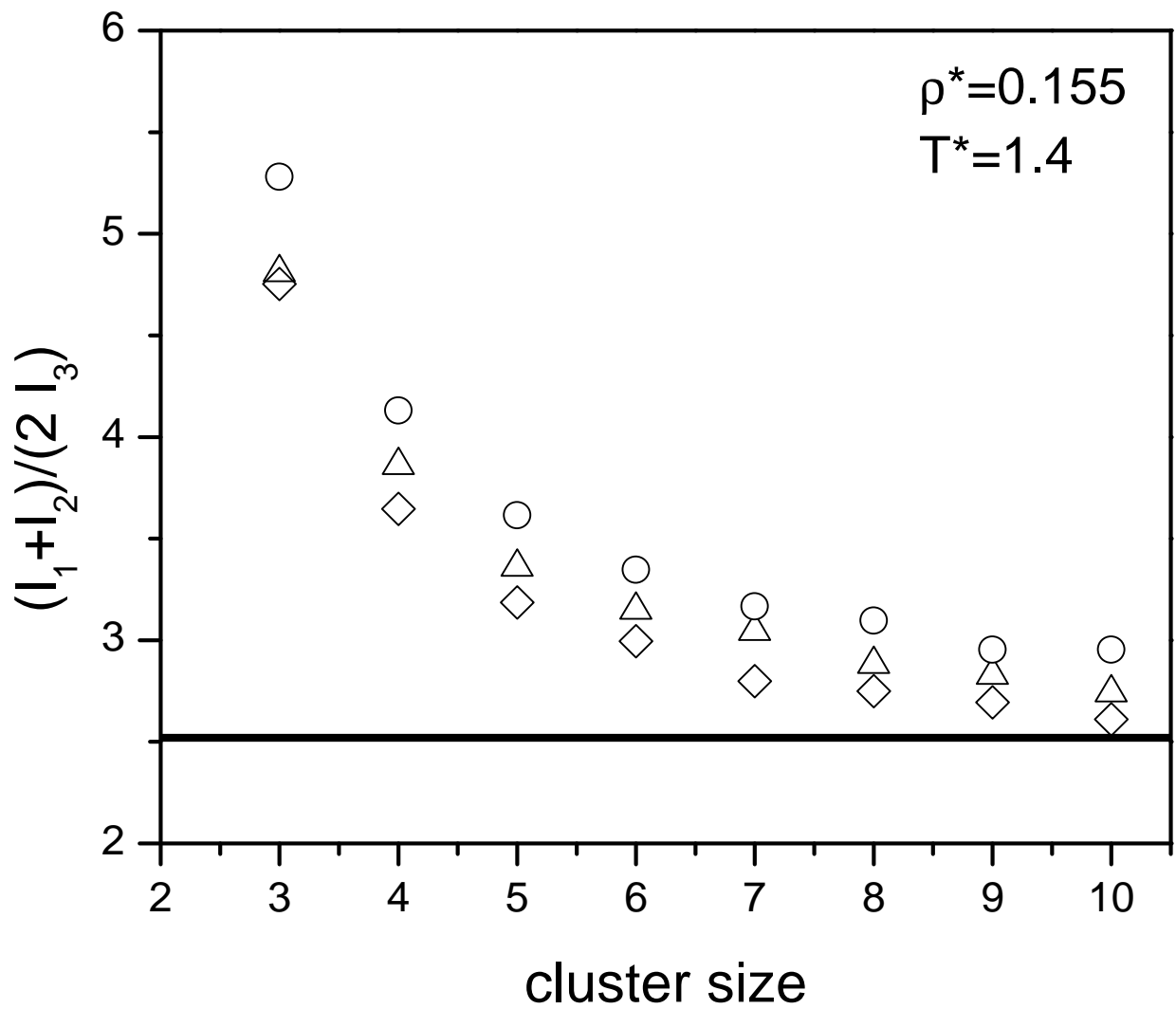
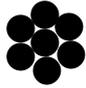
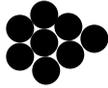
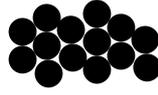
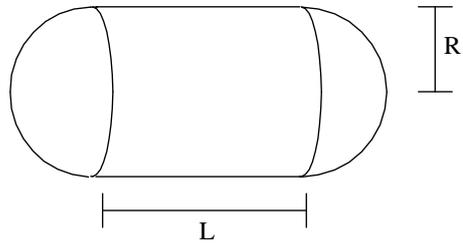
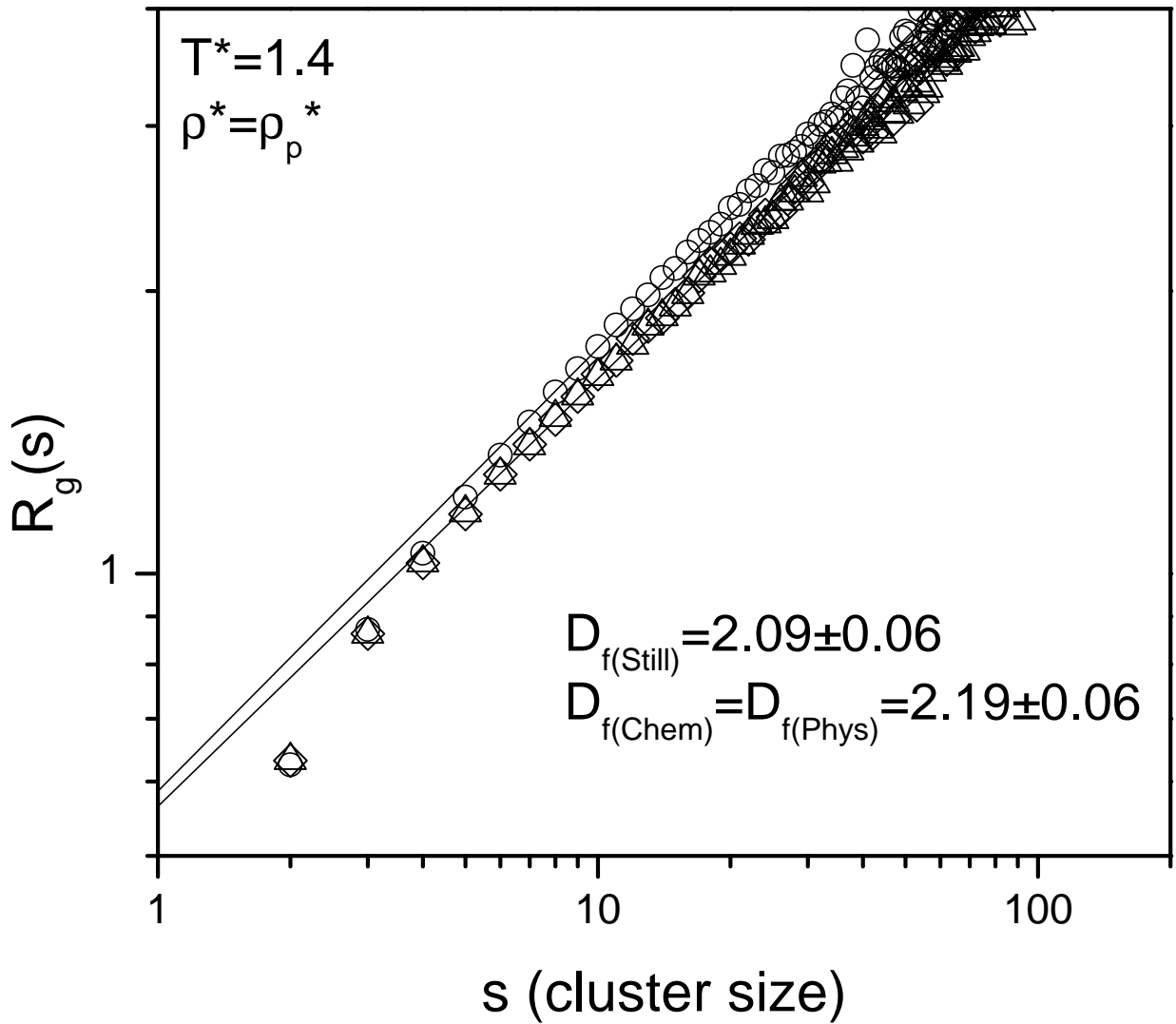
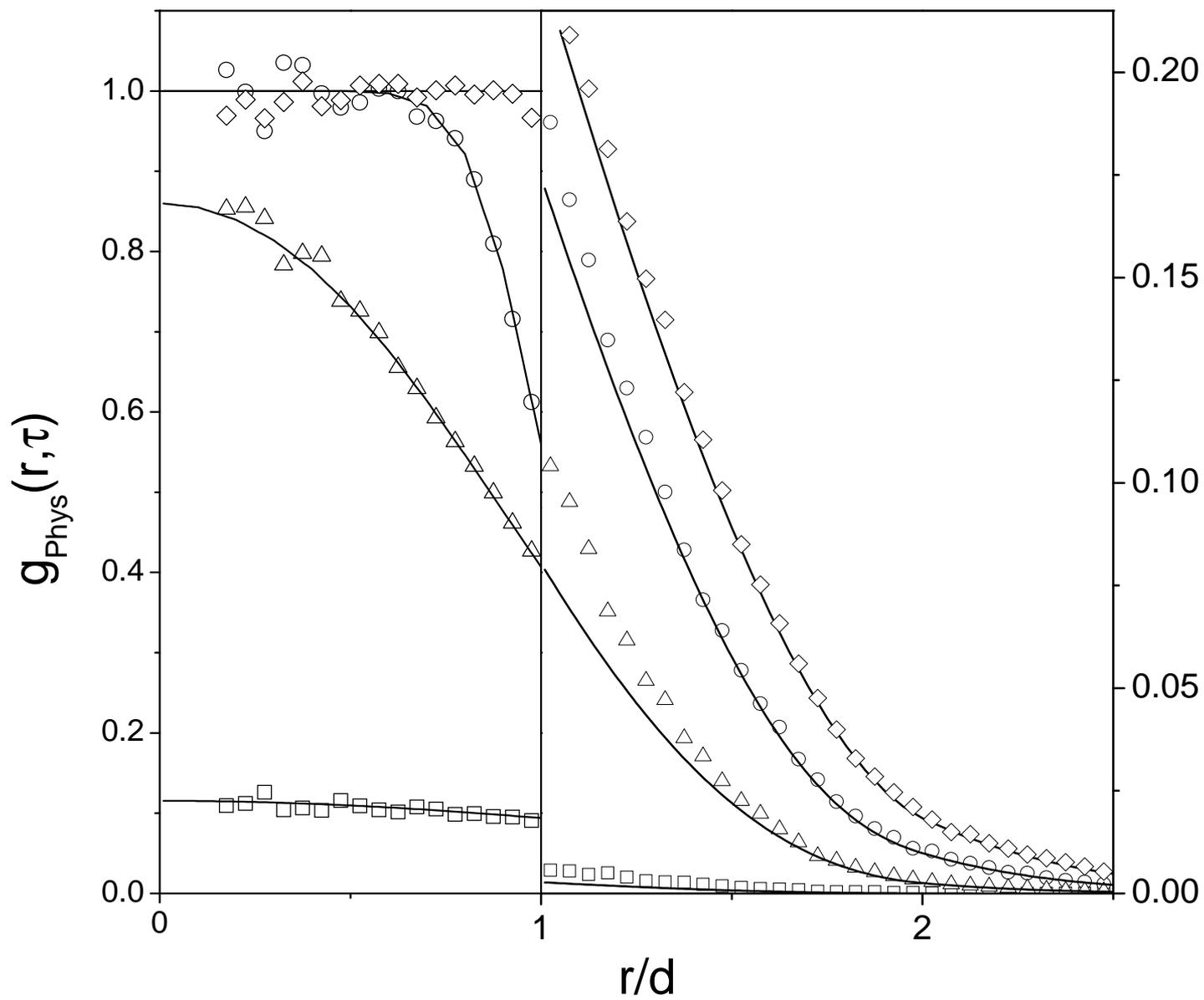
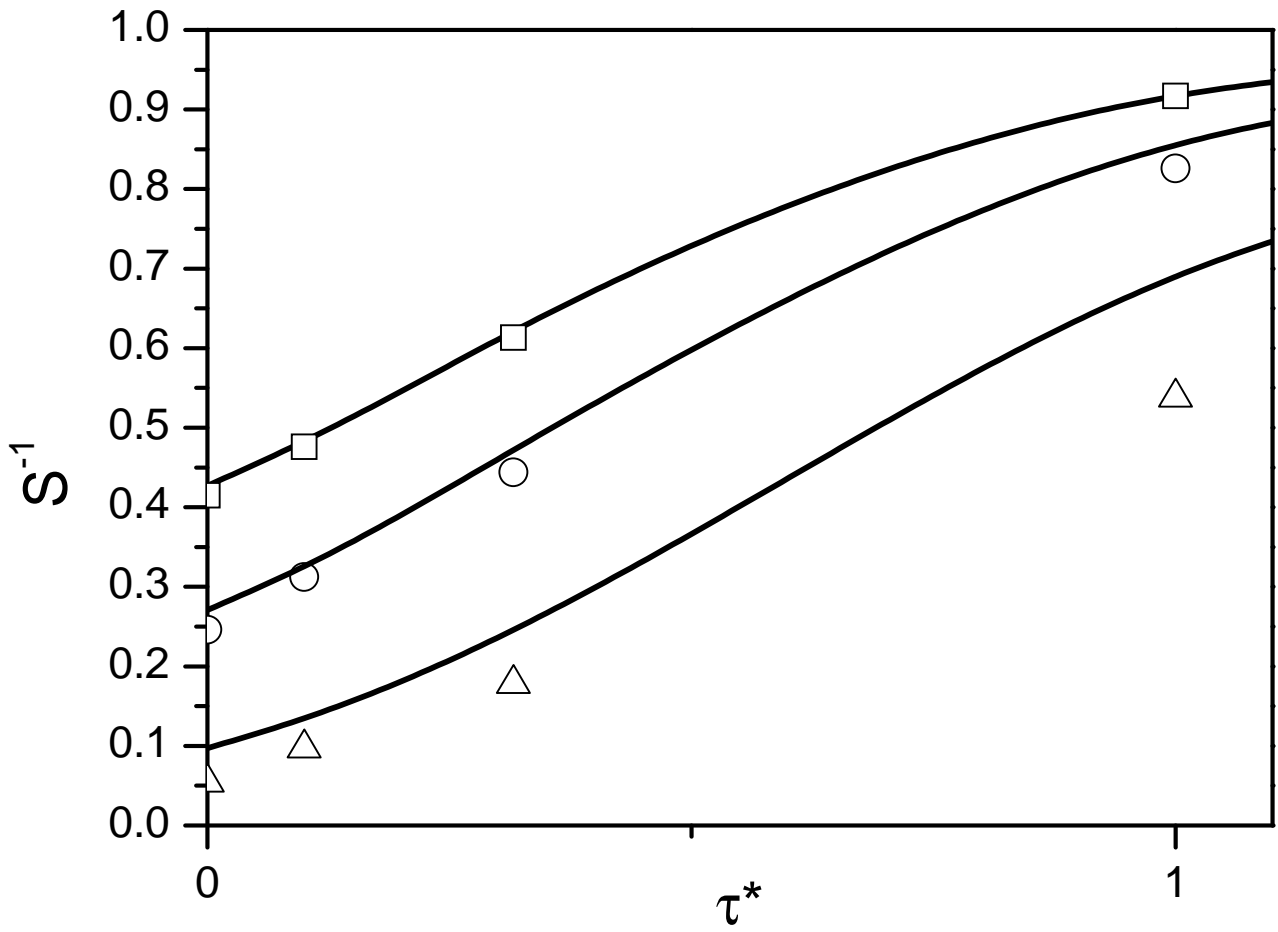
\end{document}